\documentclass[letterpaper,12pt,titlepage,oneside,final]{book}
 
\usepackage{ifthen}
\usepackage{epigraph} 
\usepackage{subcaption}
\usepackage{xcolor}
\usepackage{colortbl}
\usepackage{tabularx}
\usepackage{bbm}
\usepackage{amsmath}
\usepackage{multirow}
\usepackage{bbold}
\usepackage{adjustbox}
\usepackage{booktabs}

\newboolean{PrintVersion}
\setboolean{PrintVersion}{true} 

\ifthenelse{\boolean{PrintVersion}}{
\usepackage[top=1in,bottom=1in,left=0.75in,right=1.25in]{geometry}   
}{
\usepackage[top=1in,bottom=1in,left=0.75in,right=1.25in]{geometry}   
}

\usepackage{amsmath,amssymb,amstext} 
\usepackage{graphicx} 

\usepackage{nomentbl} 
\makenomenclature 

\usepackage{ifpdf}

\newcommand{\href}[1]{#1} 

\usepackage[\ifpdf pdftex,\fi letterpaper=true,pagebackref=false]{hyperref} 
\ifthenelse{\boolean{PrintVersion}}{   
\hypersetup{	
    citecolor=black,%
    filecolor=black,%
    linkcolor=black,%
    urlcolor=black}
}{} 

\let\origdoublepage\cleardoublepage
\newcommand{\clearemptydoublepage}{%
  \clearpage{\pagestyle{empty}\origdoublepage}}
\let\cleardoublepage\clearemptydoublepage

\begin{document}

%
\newcommand{\thesisauthor}{Ikboljon Sobirov}
\newcommand{\thesistitlecoverpage}{%
  Diagnosis and Prognosis of \\ Head and Neck Cancer Patients \\
  using Artificial Intelligence
}
\newcommand{\degree}{Master of Science} 
\newcommand{\nameofprogram}{Computer Vision}
\newcommand{\academicunit}{Department of Computer Vision}
\newcommand{\graduationyear}{2022}
%
\pagestyle{empty}
\pagenumbering{roman}

\begin{titlepage}
        \begin{center}
        \vspace*{1.0cm}

        \Huge
        {\bf \thesistitlecoverpage }

        \vspace*{1.0cm}

        \normalsize
        by \\

        \vspace*{1.0cm}

        \Large
        \thesisauthor \\

        \vspace*{3.0cm}

        \normalsize
        Thesis submitted to the\\
        Deanship of Graduate and Postdoctoral Studies\\
        In partial fulfillment of the requirements\\
        For the \degree~degree in\\
        \nameofprogram\\

        \vspace*{2.0cm}

        \academicunit\\
        Mohamed bin Zayed University of Artificial Intelligence (MBZUAI)\\

        \vspace*{1.0cm}

        \copyright~\thesisauthor, Abu Dhabi, UAE, \graduationyear\\
        \end{center}
\end{titlepage}

\pagestyle{plain}
\setcounter{page}{2}

\cleardoublepage 

%
%
\cleardoublepage 

\begin{center}\textbf{Examining Committee Membership}\end{center}
  \noindent
The following served on the Examining Committee for this thesis. The decision of the Examining Committee is by majority vote.
  \bigskip
  
  
  \noindent
\begin{tabbing}
Internal-External Member: \=  \kill 
Supervisor(s): \> Mohammad Yaqub\\
\> Professor, Department of Computer Vision,\\ 
\> Mohamed bin Zayed University of Artificial Intelligence (MBZUAI) \\
\end{tabbing}
  \bigskip
  
  \noindent
  \begin{tabbing}
Internal-External Member: \=  \kill 
Internal Member: \> Fahad Khan \\
\> Professor, Department of Computer Vision,\\ 
\> Mohamed bin Zayed University of Artificial Intelligence (MBZUAI) \\
\end{tabbing}
  \bigskip
  
  

\cleardoublepage

\begin{center}\textbf{Author's Declaration}\end{center}

  \noindent
I hereby declare that I am the sole author of this thesis. This is a true copy of the thesis, including any required final revisions, as accepted by my examiners.

  \bigskip
  
  \noindent
I understand that my thesis may be made electronically available to the public.

\cleardoublepage


\begin{center}\textbf{Abstract}\end{center}

Cancer is one of the most life-threatening diseases worldwide, and head and neck (H\&N) cancer is a prevalent type with hundreds of thousands of new cases recorded each year. Clinicians use medical imaging modalities such as computed tomography and positron emission tomography to detect the presence of a tumor, and they combine that information with clinical data for patient prognosis. The process is mostly challenging and time-consuming. Machine learning and deep learning can automate these tasks to help clinicians with highly promising results. This work studies two approaches for H\&N tumor segmentation: (i) exploration and comparison of vision transformer (ViT)-based and convolutional neural network-based models; and (ii) proposal of a novel 2D perspective to working with 3D data. Furthermore, this work proposes two new architectures for the prognosis task. An ensemble of several models predicts patient outcomes (which won the HECKTOR 2021 challenge prognosis task), and a ViT-based framework concurrently performs patient outcome prediction and tumor segmentation, which outperforms the ensemble model.

\cleardoublepage


\begin{center}\textbf{Acknowledgements}\end{center}

I would like to thank my supervisor, Dr Mohammad Yaqub for being an amazing guide, teacher and mentor, and for supporting me in pursuing the research topic of my interest. He was always extremely supportive and approachable both academically and personally, ensuring that my research work was indeed diligent and impactful. I was fully supported by the MBZUAI Scholarship program for my degree studies and research, and would like to express my gratitude for that opportunity. I was lucky enough to work with some of the most brilliant minds in AI, and my special thanks go to my esteemed friends and co-authors - Numan Saeed, Roba Al Majzoub, Otabek Nazarov, Hussain Alasmawi, and Hashmat Malik. This thesis work would not exist without the hard work, insightful ideas and continued support of these people. I also thank all the MBZUAI professors, especially Dr Fahad Khan and his team for the two excellent courses, CV701 and CV703 that immensely helped me in my research and helped me build solid foundations in the field. 

Finally, I would like to thank my parents, Sherzodbek and Ugilkhon, and the rest of my family, Islombek, Ilyosbek, Sadokatkhon, Samiyya, and Sa'diya for always believing in me, loving me unconditionally and supporting me on this journey. None of this would be possible without them. 

\cleardoublepage


\epigraph{`Actions are but by intentions, and each person will have but that which he intended.'}{\textit{Muhammad S.A.W.}}

\cleardoublepage

\renewcommand\contentsname{Table of Contents}
\tableofcontents
\cleardoublepage
\phantomsection

\addcontentsline{toc}{chapter}{List of Tables}
\listoftables
\cleardoublepage
\phantomsection		

\addcontentsline{toc}{chapter}{List of Figures}
\listoffigures
\cleardoublepage
\phantomsection		




\chapter*{List of Abbreviations}
\begin{longtable}{cp{0.8\textwidth}}

AI      & Artificial Intelligence \\
BN      & Batch Normalization \\
CNN     & Convolutional Neural Network \\
CT      & Computed Tomography \\
CoxPH   & Cox Proportional Hazard \\
DL      & Deep Learning \\
DSC     & Dice Similarity Coefficient \\
FC      & Fully Connected \\
FDG-PET & Fluorodeoxyglucose Positron Emission Tomography \\
HECKTOR & Head and Neck Tumor Segmentation and Outcome Prediction in PET/CT Images \\
H\&N    & Head and Neck \\
INHANCE & International Head and Neck Cancer Epidemiology \\
MICCAI  & Medical Image Computing and Computer Assisted Intervention \\
ML      & Machine Learning \\
MLP     & Multi Layer Perceptron \\
MRI     & Magnetic Resonance Imaging \\
MSA     & Multi-head Self Attention \\
MTLR    & Multi-task Logistic Regression \\ 
NLP     & Natural Language Processing \\
N-MTLR  & Neural Multi-task Logistic Regression \\
PET     & Positron Emission Tomography \\
ReLU    & Rectified Linear Unit \\
RT      & Radiation Therapy \\
SA      & Self Attention \\
SI      & Super Images \\
SSL     & Self-Supervised Learning \\
TNM     & Tumor-Node-Metastasis \\
UNETR   & UNet TRansformer \\
ViT     & Vision Transformer \\

\end{longtable}

\renewcommand{\nomGname}{\textbf{\large Mathematical Symbols}}

\renewcommand{\nomXname}{\textbf{\large Superscripts}}
\renewcommand{\nomZname}{\textbf{\large Subscripts}}


%

\pagenumbering{arabic}


\chapter{Introduction}

\section{Background}
Cancer is one of the leading causes of mortality worldwide. Head and neck (H\&N) cancer is the sixth most common type of cancer with 507,000 deaths in 2007~\cite{hncstats}. Clinically, radiologists perform diagnosis using medical imaging, such as computed tomography (CT) and positron emission tomography (PET) images, and if the patient has the cancer, they study electronic health records (EHR) of the patient on top of the imaging data to perform prognosis. 

The automation of tumor segmentation and patient prognosis has been studied extensively using deep learning (DL). Many solutions using convolutional neural networks (CNNs), vision transformers (ViTs), and traditional machine learning (ML) models were proposed for these tasks in various approaches. U-Net~\cite{cciccek20163d} is the most commonly used architecture with numerous variations for the segmentation task. Recently, ViT networks have been more popular in segmentation of different organs and tissues. The prognosis task is generally addressed in two ways: (i) initially extracting radiomic features from imaging data and using machine learning models for prognosis, or (ii) using an ensemble of several models concurrently extracting and producing risk scores for prognosis.

\section{Problem Statement}
Although it is one of the commonly faced cancers, the study in the automation of H\&N tumor segmentation and patient analysis using artificial intelligence (AI) models is rather limited. \textit{Head and neck tumor segmentation and outcome prediction in PET/CT images (HECKTOR)} challenge in 2020 paved the way for a further analysis into the problem. The challenge has two tasks: segmentation of H\&N tumor, and outcome prediction of patients with H\&N cancer. Still, the problem is understudied considering that the challenge is new. 

To the best of our knowledge, solutions for the segmentation of H\&N tumor heavily rely on CNNs, and no work has been carried out to see the effects of using ViT-based models for this task. Additionally, CNN models that were proposed thus far primarily make use of 3D networks for segmentation, and no 2D approach has been investigated. This shows how similar the existing solutions are for the task. Prognosis of patients with H\&N cancer in existing literature is generally performed (i) using only traditional ML models or (ii) using DL for feature extraction and then applying traditional ML models. No ensemble of models that can concurrently analyze imaging and EHR data and predict patient outcomes has been suggested. Also, ViTs, which have the ability to digest different modalities of data at the input, were not explored for prognosis.

\section{Thesis Overview}
This work studies H\&N cancer by proposing diagnostic and prognostic solutions for the problem using traditional ML and DL. Chapter~\ref{chp_literature} provides more detailed background to the problem clinically and reviews the currently existing literature on the automation of the tasks. It also mentions the gap in the research community that this work is trying to fill in. Chapter~\ref{chp_transformers} investigates the use of ViT-based model and its comparison to leading CNN networks for the segmentation task. Chapter~\ref{chp_superimages} proposes a novel 2D perspective to working with volumetric medical data and argues that this approach can be a new field of study. Chapter~\ref{chp_prognosischapters} studies outcome prediction of patients with H\&N cancer, proposing two different methods for the task. The first model is an ensemble of several DL and ML models together; and the second method is a ViT-based framework that takes the imaging and EHR data at input to perform outcome prediction and segmentation simultaneously. Finally, Chapter~\ref{chp_conclusion} concludes the thesis work and provides future directions of study for H\&N cancer. 

\chapter{Literature Review}
\label{chp_literature}

\section{Clinical Background}
\subsection{Gravity of Cancer}
Cancer is projected to become the leading cause of death worldwide in the 21st century~\cite{cancerstats}. According to The Global Burden of Disease, 890,000 new cases of and 507,000 deaths from H\&N cancer were globally reported in 2017~\cite{hncstats}, and it accounts for the sixth most prevalent cancer today~\cite{hncstatsranking}. The number of patients diagnosed with early stage is estimated to account for 30-40\%, and they are promised a five year survival rate of 70-90\% when supplemented with treatment. Unfortunately, H\&N cancer is predominantly diagnosed at later stages. Treatment is much harder with medicine and surgery at the late stages, and is highly likely to inflict serious damage to speech and swallowing functions~\cite{hashim2019head}. 

\subsection{Head and Neck Cancer}
\begin{figure}[htbp]
\captionsetup[subfigure]{justification=centering}
\centering
\begin{minipage}{0.45\textwidth}
\begin{subfigure}{\textwidth}
    \includegraphics[width=0.75\textwidth]{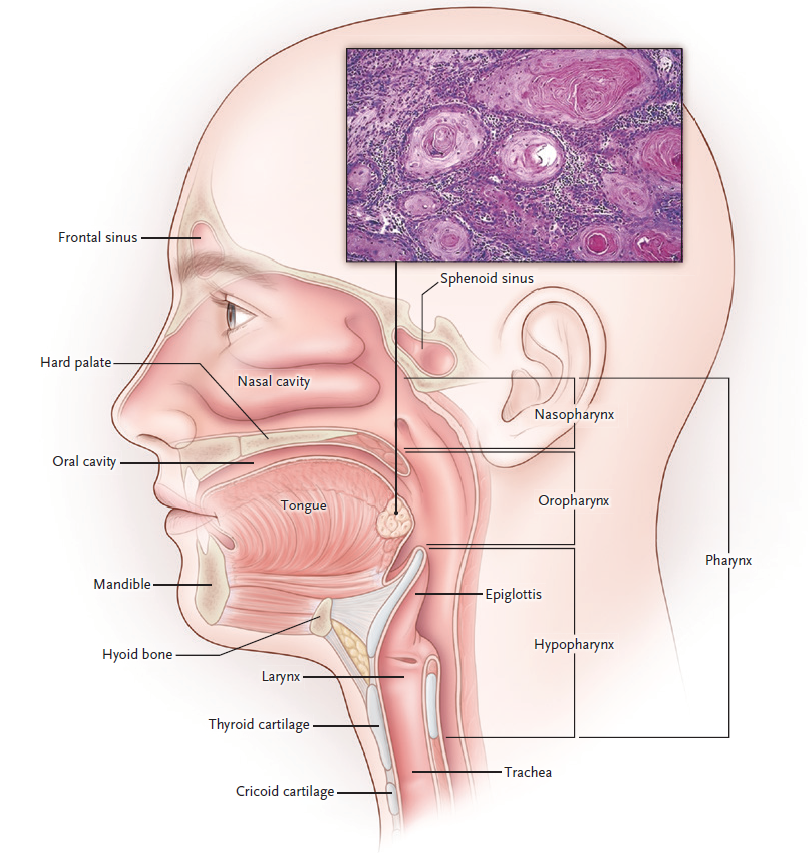}
    \subcaption{\textbf{Head and neck cancers}}
\end{subfigure}
\end{minipage}
\begin{minipage}{0.45\textwidth}
\begin{subfigure}{\textwidth}
    \includegraphics[width=\textwidth]{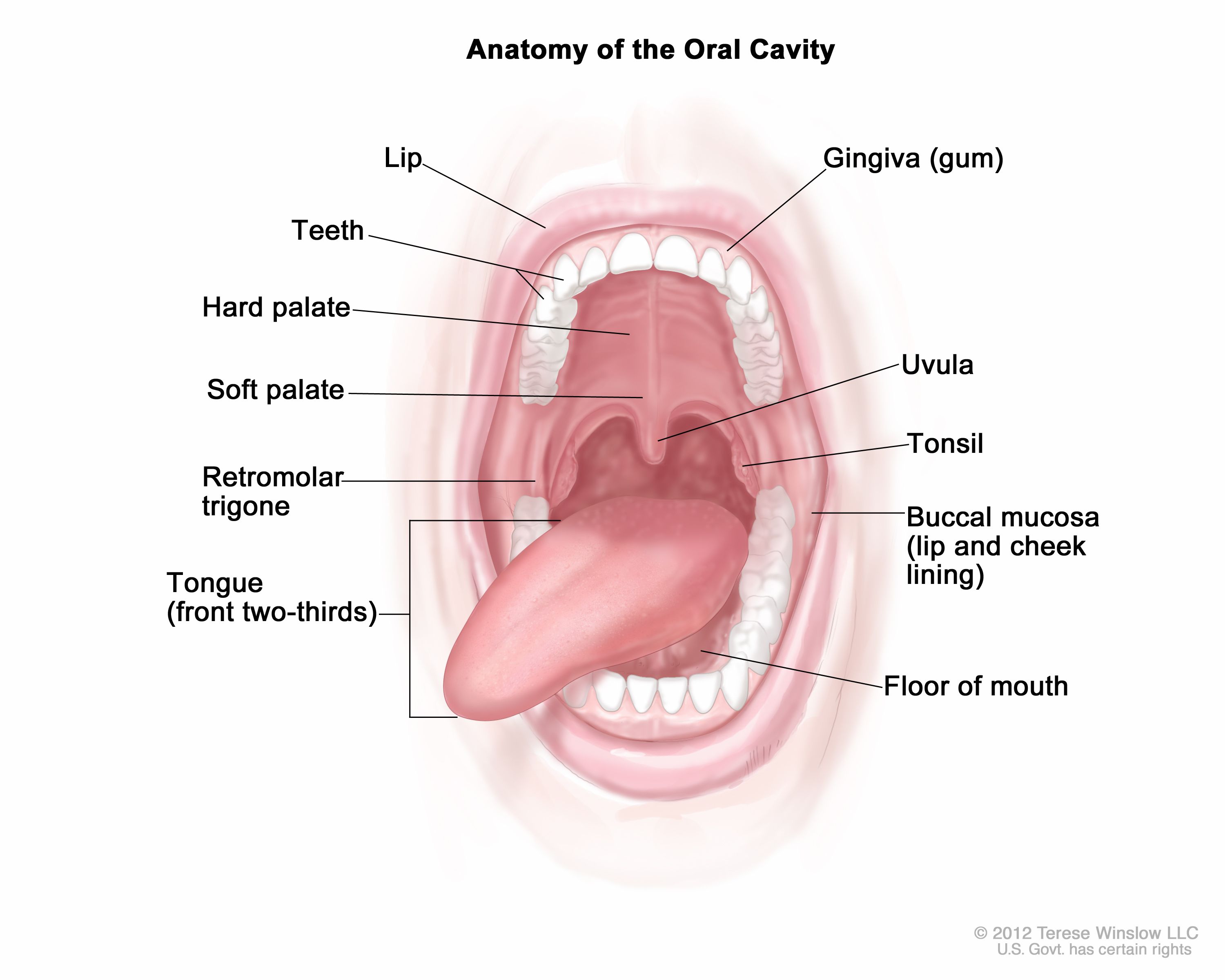}
    \subcaption{\textbf{Oral cavity}}
\end{subfigure}
\end{minipage}
\caption{Anatomical location of head and neck cancers. They arise primarily from mucosal epithelia of the oral cavity, larynx and pharynx. (a) from~\cite{chow2020head}, (b) from~\cite{winslow}}
\label{headandneckcancer}
\end{figure}

Let us start by clarifying what H\&N cancer refers to anatomically. Cancers developing in the H\&N areas are collectively referred to as squamous cell carcinomas of the head and neck. They originate in the squamous cells that form mucosal epithelia of oral cavity, oropharynx, hypopharynx, nasopharynx, and larynx as can be seen in Figure~\ref{headandneckcancer}\cite{johnson2020head}. Those that arise in sinuses, glands, nerves, or muscles in the same areas can also be framed within the same category but are much less ubiquitous. 

\subsection{Risk Factors of H\&N Cancer}
International Head and Neck Cancer Epidemiology (INHANCE) consortium was established in 2004 to conduct study on H\&N cancer in particular, including its risk factors~\cite{hncstats}. They listed tobacco and sub-products of tobacco, alcohol, human papillomavirus, occupational exposure, diet habits, oral hygiene, lifestyle, and socioeconomic level as the main contributors to the development of H\&N cancer. Amongst this list, tobacco and alcohol play the most consequential role in leading to the cancer~\cite{hncstats,johnson2020head}. In fact, heavy smokers are five to twenty-five times more prone to develop the cancer than non-smokers, and alcohol complements tobacco, increasing the risk~\cite{marur2008head}.

\subsection{Treatment of H\&N Cancer}
Until the 1980s, the only effective methods for patient curative management were surgery and radiation therapy (RT). Chemotherapy's role in designing a sequential treatment plan for patients with H\&N cancer in which induction chemotherapy is followed by RT was game-changing~\cite{marur2008head}. H\&N cancer treatment is a demanding task, requiring the presence of a radiation oncologist, a medical oncologist and a H\&N surgeon to collectively diagnose the patient with an accurate cancer stage and resectability of the tumor. The current approaches to treat patients highly depend on the tumor stage at T1, T2, T3, and T4 (for detailed differences between stages and nodes, refer to~\cite{chow2020head}). T1 and T2 stage patients, defined as early stage, do not yet develop nodes, and are thus easier to treat with surgery or RT. Concurrent chemoradiation (i.e. induction chemotherapy followed by RT) is applied as treatment to more advanced tumors~\cite{marur2008head}. As can be seen, the role of RT in treating patients with all H\&N cancer types is substantial, with each patient receiving RT at least once during the course of treatment~\cite{strojan2017treatment}.

\subsection{Diagnosis of H\&N Cancer}
Once we understand how significant this disease is and how we can treat cancerous patients, let us step back and go over the diagnosis and prognosis procedures for H\&N cancer patients. Once a patient is admitted to a hospital, a detailed history is collected and a physical examination is carried out by clinicians. Radiographic imaging is required to detect the presence and location of a tumor's region, and is proven to be much safer than biopsy especially when diagnostic analysis is incomplete~\cite{chow2020head}. 

Radiographic imaging is used as one of the initial stages for cancer diagnosis since it is a non-invasive and inexpensive approach compared to biopsy sampling via surgery, for example. Primarily, MRI, CT and PET are the three prevalent types of modalities utilized for imaging head and neck areas~\cite{marur2008head}. Note here that all three modalities are designed to capture volumetric image of the targeted organ. The innate nature of tumor also requires volumetric imaging capacity. It is substantially beneficial for doctors to accurately diagnose and prognose a patient using the tumor volume computed by the 3D data~\cite{ahmad2014medical}. 

Endoscopy and biopsy are performed once an abnormality is detected via physical examination; these two can come before or after the radiographic imaging. It is preferred to capture radiography before endoscopy and biopsy since radiography is much faster, non-invasive, less dangerous and cheaper~\cite{marur2008head}. Imaging can be used after them to accurately localize the tumor and use the volumetric data for tumor staging. Imaging also helps with understanding the metastases that may be spreading over other neighboring organs~\cite{ha2006role}. 

CT has the capability to provide structural formation of the organs of interest, whereas PET can easily highlight metabolically active tissues such as the tumor~\cite{ha2006role,marur2008head}. This has led to the application of the two modalities in parallel to diagnose head and neck cancer. Clinicians can effortlessly pinpoint the locality and the extent of tumor using PET and understand its structural location using CT; and therefore, the conjunction of the two modalities are the most prevalent these days~\cite{ha2006role}. In this thesis work as well, PET and CT scans of patients diagnosed with H\&N cancer are used in combination.

\subsection{Prognosis of H\&N Cancer}
\label{prognosis_of_hnc}
Once clinicians identify if a patient is suffering from H\&N tumor, and if they are, how serious the disease is, the next most essential step would be to do prognosis. With accurate prognosis, doctors can plan the right course of treatment, may it be chemotherapy, RT, surgery, or the combination of them. The foremost prognostic assessment tool is the tumor-node-metastasis (TNM) staging method~\cite{sobin2011tnm}. This scheme assesses the patient to be classified into rankings using the primary tumor (T), regional lymph nodes (N), and distant metastasis (M). Such schemes as TNM are stored in EHR for each patient to perform prognosis. 



\section{Automation of Prognosis and Diagnosis}

\subsection{Literature Review on Prognosis}
\label{ensemble_models}
Before the prevalence of DL, traditional ML and statistical models were mostly used for the automation of the prognosis task. Chapter~\ref{chp_prognosischapters} focuses on two distinct solutions for the prognosis of patients with H\&N cancer, and as such this section summarizes the literature that performed the task using statistical, ML and more recent DL approaches. 

\subsubsection{Statistical Models}
Statistical models were the foundation for automatic prognosis of patients with different types of diseases. One of the earliest, yet still commonly used, approaches is Cox proportional hazard (CoxPH) model proposed by Cox in 1972~\cite{cox1972regression}. This approach models a survival (or hazard) function that produces probability of a particular event occurring at a given time $t$. It is widely used in the medical research to predict the survival time of patients based on a set of given clinical features. The hazard function $h(t, x_i)$ assumes that time element $\lambda_0(t)$ and feature element $\eta(\overrightarrow{x_i})$ are proportional, and is defined as~\cite{cox1972regression}:
\begin{equation}
    h(t, \overrightarrow{x_i}) = h_0(t)\eta(\overrightarrow{x_i})
\end{equation}
where $h_0(t)$ is the baseline reliability function (generally not specified), and $\eta(\overrightarrow{x_i})$ is the risk function. The risk function $\eta(\overrightarrow{x_i})$ is often framed with a linear representation such that $\eta(\overrightarrow{x_i}) = exp(\sum_{j=1}^{p}x_j^i\omega_j) . \omega_j$ are the coefficients to find~\cite{kumar1994proportional,bender2005generating}. The equation is adapted from~\cite{pysurvival_cite}.

An alternative to CoxPH is the MTLR model that can essentially be described as a sequence of logistic regression models dedicated to a number of time intervals whose task is to calculate the probability of an event occurring in that specific time interval. Yu~\cite{yu2011learning} developed this model in 2011 to tackle the issues introduced by CoxPH counterpart. Specifically, CoxPH brings three major issues: (i) the time component of the hazard function is unspecified, which makes the model unfitting to actual risk prediction, (ii) it assumes that the ratio of hazards for two given patients is constant over time, and (iii) the function is not computationally efficient when it concerns the formula responsible for ties, forcing the model to use approximations~\cite{yu2011learning,pysurvival_cite}. The MTLR model can handle these issues. Its loss function is defined as:

\begin{equation}
    \min _{\Theta} \frac{C}{2} \sum_{j=1}^{m}\left\|\vec{\theta}_{j}\right\|^{2}-\sum_{i=1}^{n}\left[\sum_{j=1}^{m} y_{j}\left(s_{i}\right)\left(\vec{\theta}_{j} \cdot \vec{x}_{i}+b_{j}\right)-\log \sum_{k=0}^{m} \exp f_{\Theta}\left(\vec{x}_{i}, k\right)\right]
\end{equation}

\normalsize
where the smoothness of the predicted survival curves depends on the change between consecutive time intervals and is controlled by the constant $C$ (set as a hyperparameter) in the $l_2$ regularization term. 

Fotso~\cite{fotso2018deep} in 2018 integrated neural networks on top of MTRL (dubbing as N-MTLR) to introduce nonlinearity to the model. In other words, when there is nonlinear elements in data, the two previous models fail to produce satisfactory results since they both are based on a linear transformation. Neural nets introduced in the N-MTLR solves this issue, outperforming the MTLR model in numerous applications and CoxPH when nonlinearity is found in data~\cite{fotso2018deep}.     

\subsubsection{ML \& DL Models}
While statistical models make good use of data when making decisions, they do not directly craft useful features from them. ML, on the other hand, is able to automatically recognize patterns in data, thus acquiring actionable and meaningful knowledge~\cite{goodfellow2016deep}. Computers' capability to learn from data to make well-informed decisions in intelligent heathcare systems started to gain more popularity with the machine learning tools and algorithms getting more diverse and prevalent. ML has advanced the automation of numerous tasks of human specialists, yielding real and tangible results. Therefore, an extensive amount of research work has been put into the automation of prognosis of patients using their clinical data with machine learning algorithms. The patient prognosis of breast~\cite{yue2018machine,ganggayah2019predicting,park2013robust}, oral ~\cite{chang2013oral,chuang2010support,hung2020artificial}, bladder~\cite{song2020machine,hasnain2019machine}, prostate~\cite{zupan2000machine,hou2018rankprod}, and lung~\cite{lynch2017fuqua,sim2019predicting} cancer are some of the examples for their acclaim in the research community.

Along with the clinical information of patients diagnosed with cancer, most studies attempt to integrate meaningful features extracted from the imaging data. \textit{Radiomics} refers to the automatic extraction of quantitative mineable features from radiographic images, such as CT, MRI and PET, that can be helpful in survival prediction and other predictive tools~\cite{yang2019development,giraud2019radiomics}. Such features can be characterised by tumor image intensity, shape, texture, multiscale wavelet and so forth, and a high number of them need to be extracted and filtered carefully to keep the meaningful and useful ones. Many papers on the prognosis of H\&N cancer patients as well make use of radiomic features in conjunction with the clinical data~\cite{wong2016radiomics}.  

Howard et al.~\cite{howard2020machine} compared three different ML survival algorithms to study whether intermediate-risk H\&N cancer patients would benefit from adjuvant chemotherapy. They used 33,527 patient cases from National Cancer Database~\cite{bilimoria2008national} (the largest clinical cancer registry) to construct DeepSurv~\cite{katzman2018deepsurv}, random survival forests~\cite{ishwaran2008random} and N-MTLR~\cite{fotso2018deep} models for the patient survival prediction for treatment recommendations with C-index of 0.693, 0.695 and 0.691 respectively. 

Parmar et al.~\cite{parmar2015radiomic} studied the prediction of overall survival of H\&N cancer patients. They extracted 440 radiomic features from two cohorts of H\&N cancer patient CT scans using 14 feature selection methods. They investigated 12 ML classifiers to compare the prognostic performance and stability of these models and the feature selection methods using the assessment with the area under receiver operator characteristic curve (AUC). Random forests (AUC=0.67), Bayesian (AUC=0.67) and nearest neighbors (AUC=0.62) were the best performing classifiers for the prognosis, and minimum redundancy maximum relevance (AUC=0.69), conditional infomax feature extraction (AUC=0.68) and mutual information feature selection (AUC=0.66) were found to yield the highest prognostic performance. 

Kazmierski et al.~\cite{kazmierski2021machine} organized a challenge for the survival analysis of patients with H\&N cancer. The primary goal of the challenge was to predict two-year overall survival, and predicting death risk of a patient and full survival curve was the secondary task. The data included EHR (demographic, clinical, and interventional) and pre-treatment contrast-enhanced CT scans. Twelve submissions were received in the challenge that use EHR data only, imaging data only, or fuse them in various fashions. Participants who used the imaging data mostly relied on DL to extract features automatically. Four submissions used the hand-crafting approach to extract representations from the imaging data. Their observation over the submissions shows that the most models proposed do not learn valuable representations from radiomic features even if they are considered to be the common understanding, claiming that the EHR are the primary source of model predictions. The authors signify the importance of tumor volume as a simple yet powerful predictor for the prognosis. The most intriguing aspect of the challenge is that, even if there were many DL-based model submissions, the winning solution used Deep MTLR model using EHR data and tumor volume for the task. 




\subsection{Literature Review on Diagnosis}

\subsubsection{H\&N cancer diagnosis}
The automatic diagnosis of different cancer types is well-researched~\cite{munir2019cancer}. Breast~\cite{qi2020automated,khuriwal2018breast}, oral~\cite{ariji2019contrast}, and brain~\cite{havaei2017brain} are some of the examples for such studies. However, research on H\&N tumor segmentation using deep learning is limited. The study on the automatic diagnosis of H\&N cancer has gained more popularity after the introduction of the HECKTOR challenge, with 18 teams participating in 2020 and 44 teams in 2021~\cite{andrearczyk2022overview}. Chapter~\ref{chp_transformers} and~\ref{chp_superimages} study the automatic segmentation of H\&N tumor using different DL approaches, and as such this section looks over the recent solutions proposed for the diagnosis task using CNN approaches. 

Iantsen et al.~\cite{iantsen2020squeeze} used squeeze and excitation normalization (SE norm) layers on top of the traditional 3D U-Net~\cite{cciccek20163d} architecture with residual blocks. Their model achieved DSC of 0.759 in the 2020 HECKTOR challenge testing set, winning the competition. The SE norm is similar
to instance normalization~\cite{ulyanov2016instance} but different in shift and scale values, which
are treated as functions of input \textit{X} during inference. SE norm is used in the encoder within the residual blocks and in the decoder after convolutional blocks. They used an ensemble of the same model for different splits to reach such a performance.

Hybrid active contour was integrated within the U-Net model in the work by Ma and Yang~\cite{ma2020combining}. This traditional ML technique, combined with the powerful CNN model, saved them the second place in the challenge. Such an additional integration was performed by Yua~\cite{yuan2020automatic} who implemented a dynamic scale attention network on top of U-Net for the segmentation task. They claim that this approach helps enhance the utilization of feature maps coming from the encoder to the decoder. The scale attention network combines different scale features using a scale attention block for each decoder layer that is connected to all the extracted features (except the last encoder layer). They prove their model performs better than the vanilla U-Net.

In~\cite{xie2021head}, Xie and Peng used a similar architecture to~\cite{iantsen2020squeeze}, a U-Net model with SE norm, with a difference in the learning rate being controlled with polyLR. They split the data into five-folds for training and validation, and five test predictions were then ensembled for the final predictions. They won the 2021 challenge with DSC of 0.7785. 

An et al.~\cite{an2021coarse} implemented three subsequent U-Net models as a single framework, each model being responsible for a different task. The first U-Net is utilized to coarsely delineate the tumor region and select the bounding box. The second model is responsible for a finer segmentation output using the small bounding box. The final model then takes the PET and CT concatenation and the preceding segmentation output for the final refined predictions. They achieved DSC of 0.7733, saving the second spot in the 2021 competition.

\subsubsection{Transformers for diagnosis}
All the solutions proposed in both the 2020 and 2021 challenges heavily relied on CNN architecture, namely U-Net with certain variations and ensembles. However, the recent years have witnessed a growing interest in transformer-based architectures~\cite{xu2021levit,wang2021transbts,xie2021cotr,zhang2021transfuse,hatamizadeh2022unetr}. Chapter~\ref{chp_transformers} studies the use of transformers for the H\&N tumor segmentation, and as such, this section briefly discusses the recent advances in ViT-driven architectures for volumetric segmentation.

The use of transformers in the segmentation task still requires a U-shaped architecture designs, and the ViT can be used in the encoder, decoder, or the both. For example, Wang et al.~\cite{wang2021transbts} effectively encoded local and global representations in depth as well as spatial dimensions to perform brain tumor segmentation. Spatial feature extraction was performed using a 3D CNN encoder. Following that, transformers were used for the global context, and finally a 3D CNN decoder to upsample back to the original dimensions. In~\cite{zhu2021region}, Zhu et al. integrated region awareness into transformers to segment breast tumor. They proved the model outperforms CNN-based counterparts using extensive experiments on ultrasound breast scans.  

UNETR model, proposed in~\cite{hatamizadeh2022unetr}, applies 12 layers of ViT in the encoder that generates features at different layers and links them back in the decoder as skip connections. The model is tested on brain tumor segmentation and abdominal multi-organ segmentation tasks, and the authors achieve comparable results to other leading models. 

To the best of our knowledge, the proposed solutions for H\&N cancer diagnosis all apply CNN-based U-Net variations, and no work has been carried out to see the performance of transformers for the task. The study on the use of transformers for the H\&N cancer segmentation and its performance comparison to other leading CNN models is given in Chapter~\ref{chp_transformers}.





\section{Metrics of Performance}
In this section, the metrics of performance reported in this thesis work are provided and explained. For the prognosis task, C-index is used as the measure of risk scores, and for the segmentation task, dice similarity coefficient (DSC), precision and recall are calculated to measure the robustness of the proposed models. 
\subsubsection{C-index}
The concordance index is a generalization of the area under the receiver operating characteristic curve that takes censored data into consideration and is defined as:

\begin{equation}
    C-index = \frac{\sum_{i,j}\mathbbm{1}_{T_i>T_j} \cdot \mathbbm{1}_{\eta_i>\eta_j} \cdot \delta_j}{\sum_{i,j}\mathbbm{1}_{T_i>T_j} \cdot \delta_j}
\end{equation}

where:
\begin{itemize}
    \item $\eta_i$ is the risk score of a unit $i$,
    \item $\mathbbm{1}_{T_i>T_j} =  
        \begin{cases}
            1 & \text{if } T_i>T_j \\
            0 & \text{otherwise}
        \end{cases}$, and $\mathbbm{1}_{\eta_i>\eta_j} =  
        \begin{cases}
            1 & \text{if } \eta_i>\eta_j \\
            0 & \text{otherwise}
        \end{cases}$
\end{itemize}

C-index of 1.0 represents that the model achieved its most optimal performance, and C-index of 0.5 means that the model is randomly predicting. The equation is adopted from~\cite{fotso2018deep}. C-index is used in all prognostic models' evaluations.

\subsubsection{Dice Similarity Coefficient}
Dice similarity coefficient (DSC) measures the spatial overlap of the two volumes or areas within a range of 0 to 1, where 0 represents no overlap between the two sets of binary segmentation results and 1 represents a full overlap. It was first proposed by Dice~\cite{dice1945measures}.

\begin{equation}
    Dice(A, B) = \frac{2|A\cdot B|}{|A| + |B|}
\end{equation}
where $A$ and $B$ are the two binary vectors; one represent the ground truth values and the other represent the prediction values. In lucid words, the score is calculated as two times the overlap divided by the sum of the two volumes/areas. DSC is used in all segmentation models' evaluations.

\subsubsection{Recall and Precision}
To understand recall and precision, a confusion table is generally used. To illustrate it, let us take the following scenario: there is a binary label $l$ with which each object in the dataset is linked, and it represents the ground truth; there is a prediction $z$ which represents the model output for the given input, as shown in Table~\ref{confusion_matrix}. $TP$, $FN$, $FP$, and $TN$ indicate true positive, false negative, false positive, and true negative values; + and - indicate the positive and negative values respectively. Equations are adopted from~\cite{goutte2005probabilistic}. 

\begin{table}[htbp]
\caption{Confusion matrix where $l$ is the ground truth label and $z$ is the model prediction.}
\label{confusion_matrix}
\begin{center}
\begin{tabular}{cc|cc}
    \arrayrulecolor{gray}\hline
    \multicolumn{2}{c}{\multirow{2}{*}{}}&\multicolumn{2}{c}{Prediction $z$}\\
    \multicolumn{2}{c}{}& + & -\\
    \arrayrulecolor{gray}\hline
    \multirow{2}{*}{Label $l$}& + & $TP$ & $FN$\\

     & - & $FP$ & $TN$\\
    \arrayrulecolor{gray}\hline
\end{tabular}
\end{center}
\label{tab:multicol}
\end{table}

Using this confusion matrix, we now can calculate recall and precision measures as follows:

\begin{equation}
    Recall = \frac{TP}{TP + FN}
\end{equation}

\begin{equation}
    Precision = \frac{TP}{TP + FP}
\end{equation}

Intuitively, recall can be understood as the measure of how many instances were correctly predicted of all the actual positives, whereas precision is the measure of how many instances were correctly predicted of the all predicted values. Precision and recall are used in segmentation models' evaluations.

\section{Research Gap}


Now that we have some background on the criticality of H\&N cancer and how much work has been carried out for the automation of the tumor segmentation and patient prognosis, this section summarizes the research gap that the thesis work addresses. 

H\&N tumor segmentation studies primarily utilize U-Net and variations of it based on CNNs. All the proposed solutions for the HECKTOR challenge in 2020 and 2021 heavily rely on CNNs for their work~\cite{andrearczyk2022overview}. However, the introduction of ViT models were game-changing with its many applications in medical imaging tasks regarding other problems. Chapter~\ref{chp_transformers} studies the use of transformers for this problem and makes a comparison of its results to other leading CNNs.

Chapter~\ref{chp_superimages} introduces a novel approach to working with volumetric data in a 2D fashion by casting the data to super images. This work contributes to the study of 3D versus 2D networks for 3D medical images, which is a heated discussion. It argues that generating super images and using 2D networks, which are much more lightweight, can perform similar to 3D networks. 

Finally, Chapter~\ref{chp_prognosischapters} fills in the gap of prognosis of patients diagnosed with H\&N cancer by incorporating clinical data with imaging data in two different ways. The first part ensembles the three different models to make risk predictions. Being similar to previous works, this approach not only outperforms other models, but also introduces a new way to integrate EHR and volumetric imaging data. The second part introduces a novel method for blending EHR and imaging data at the input level with the use of transformers. 


\chapter{Automatic Segmentation of Head and Neck Tumor: How Powerful Transformers Are?}
\label{chp_transformers}

\section{Introduction}

Globally, cancer is considered to be one of the leading causes of death, and H\&N cancer is one of the most commonly encountered types. Clinically, PET and CT are used in conjunction to detect, segment, quantify and stage the tumor region, but it is highly time-consuming, extensively labor-demanding and prone to error. ML and DL models have been proposed to automatically segment the tumor region that yield comparable results to the result of a clinician. However, most of them rely on using CNN networks for the segmentation of H\&N tumor. This chapter investigates a vision transformer-driven model to delineate H\&N tumor and compare its results to dominant CNN-based counterparts. To do that, we use the imaging data of CT and PET from the HECKTOR challenge~\cite{oreiller2022head,andrearczyk2022overview}. The transformer-based network shows the potential to yield results comparable to CNN models. It achieves a mean DSC of 0.736, mean precision of 0.766 and mean recall of 0.766 cross validated in-house; compared to that, the 2020 challenge winning model in the same cross validation achieves merely 0.021 more. On the challenge testing set, the transformer-based model achieves 0.736 DSC, 0.773 precision and 0.760 recall, being only 0.023 lower in DSC than the 2020 challenge winning model. We show that the use of transformers in the cancer segmentation task is a promising research area. 


With recent strides in DL, automatic segmentation of tumor found in various body parts is being explored with great interest. The main driver to using DL models as an alternative or auxiliary to radiologists in the task is that the former can perform as well as the latter in most cases, with an additional benefit of speeding up processing time. 

Although the prevalence of H\&N cancer is high worldwide, its study on automation using DL is insufficient in literature. The HECKTOR challenge~\cite{oreiller2022head,andrearczyk2022overview} has boosted the study of the diagnosis and prognosis of H\&N cancer using DL. At a glance, most of the proposed solutions in the segmentation task of the competition relied on U-Net~\cite{ronneberger2015u} and its variations with additional techniques and features such as 3D convolution blocks~\cite{cciccek20163d}, squeeze and excitation blocks~\cite{hu2018squeeze}, residual blocks~\cite{he2016deep,zhang2018road}, multi-scale patches~\cite{li2017multi}, so forth. For example, Iantsen et al.~\cite{iantsen2020squeeze} used squeeze and excitation normalization on residual 3D U-Net; Ma and Yang~\cite{ma2020combining} integrated 3D U-Net with hybrid active contour to perform the task; and Yuan~\cite{yuan2020automatic} used a dynamic scale attention network within 3D U-Net for the segmentation task. 

Transformers, in tandem with CNNs, are gaining more popularity in the community, with their use in the segmentation task being no exception. Numerous variations of ViT~\cite{dosovitskiy2020image}-based models were proposed for the tumor segmentation~\cite{xu2021levit,wang2021transbts,xie2021cotr,chen2021transunet,zhang2021transfuse,hatamizadeh2022unetr}. However, to the best of our knowledge, no work has been carried out in analyzing transformer-based models in H\&N cancer diagnosis. Transformer architectures are claimed to be superior to CNNs in capturing long-range dependencies~\cite{khan2021transformers}, and understanding the context information is assumed to benefit the network to accurately segment tumor regions. Furthermore, transformers are relatively new and understudied compared to CNNs, thus exploring their use in various applications would further benefit other tasks. Contributions of this work are as follows:

\begin{itemize}
    \item Investigating the comparison of a transformer-based model and leading CNN based counterparts;
    \item Showing that the transformer model can perform as well as the CNN models;
    \item Showing that data augmentations are essential to the transformer model's performance;
    \item Studying a multimodal setting of CT and PET in the case of transformers;
    \item Testing the validity of a newly proposed network in a new medical task.
\end{itemize}

\section{Methodology}

\begin{figure}[htbp]
    \centering
    \includegraphics[width=1\textwidth]{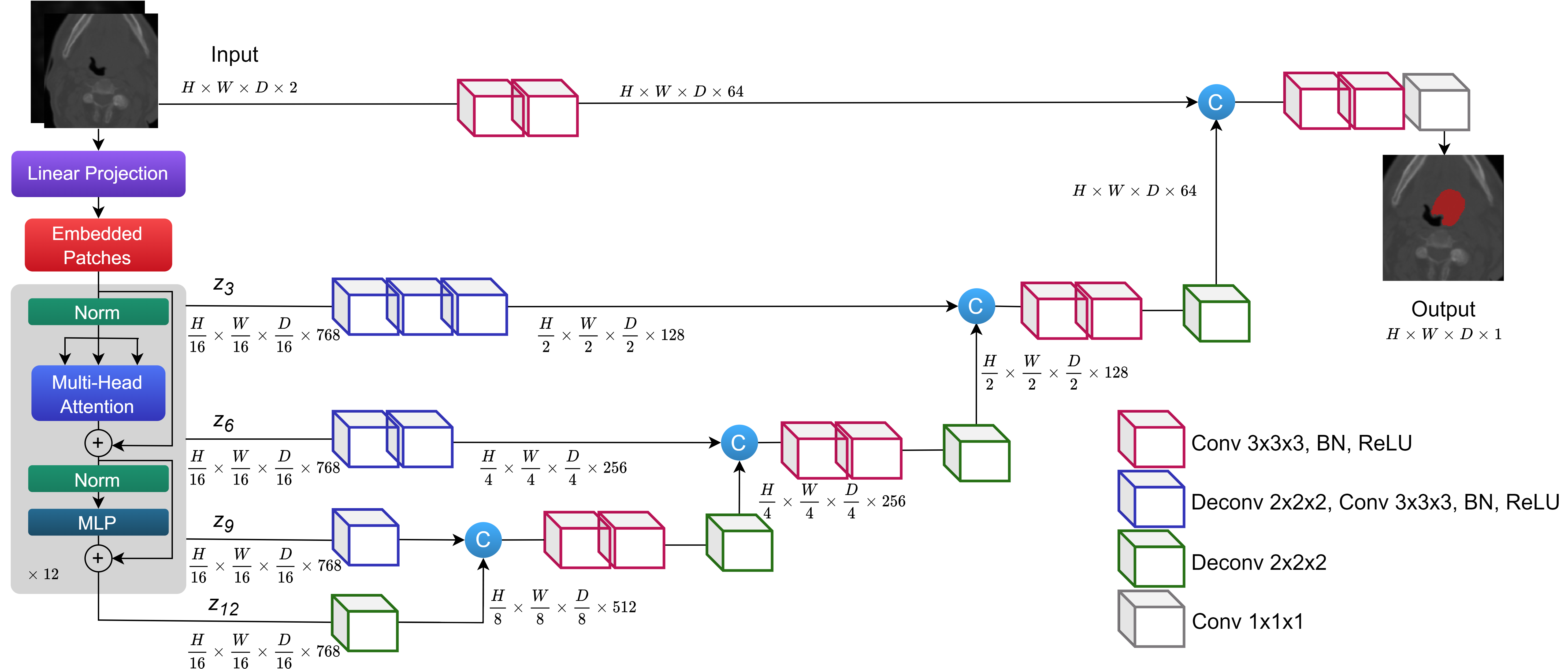}
    \caption{Overall architecture of UNETR. Features are extracted using a ViT encoder; CNN deconvolutional and convolutional blocks are used as a decoder; skip connections are used between the encoder and the decoder.}
    \label{fig:network_unetr}
\end{figure}

\subsection{Architecture}
Since their success in NLP, transformers have been welcomed into the computer vision tasks as well, with ViT~\cite{dosovitskiy2020image} laying the groundwork. Inspired by Hatamizadeh et al.~\cite{hatamizadeh2022unetr}, we developed a transformer-based model for H\&N tumor segmentation and compared its results to best performing CNN models. The network is depicted in Figure~\ref{fig:network_unetr}. The overall architecture composes of an encoder, a decoder and skip connections between the two, mimicking the standard U-Net design. 

A 3D volumetric input has the size of $x \in \mathbb{R}^{H\times W\times D\times C}$, where \textit{H, W, D} denotes the height, width and depth respectively, and $C$ is the number of channels. It gets embedded into a 1D sequence of flattened uniform non-overlapping patches with a new size of $x_{v} \in \mathbb{R}^{N\times (P^3.C)}$, where $N=(H\times W\times D)/P^3$ is the length of the sequence, and $P^3$ is the resolution of each patch.

Following the patch embeddings, a linear layer is used to project them into $K$ dimensional latent space (remains constant). An additional 1D learnable positional encoding $E_{pos} \in \mathbb{R}^{N\times K}$ is attached to the projected patch embeddings $E\in \mathbb{R}^{(P^3.C)\times K}$ to keep the spatial information as:

\begin{equation}
    z_0 = [x_v^1E; x_v^2E;\dots;x_v^NE]+E_{pos}
\end{equation}

Twelve layers of transformer blocks, with multi-head attention (MSA) and multi-layer perceptron (MLP), follow the embedding layer. MSA and MLP follows as such:

\begin{equation}
    z_i^{\prime} = MSA(Norm(z_{i-1}))+z_{i-1}, \quad\quad i=1\dots L,
\end{equation}

\begin{equation}
    z_i = MLP(Norm(z_i^{\prime}))+z_i^{\prime}, \quad\quad i=1\dots L,
\end{equation}
where $Norm()$ is a layer normalization as in~\cite{ba2016layer}, $i$ is the identifier in the intermediate block, $L$ denotes the number of transformer blocks, and $MLP$ consists of two linear layers.

An $n$ number of self-attention (SA) heads, existent in the $MSA$ component, learn mapping between a $q$ query and the respective $k$ key and $v$ value in a given sequence of $z\in \mathbb{R}^{N\times K}$. The SA for the given sequence $z$ is calculated as the following:

\begin{equation}
    SA(z) = (\sigma(\frac{qk^T}{\sqrt{K_h}}))v,
\end{equation}
where $\sigma()$ is the softmax function, $K_h = K/n$ is the scaling factor, and $v$ is the values in the input sequence. Respectively, $MSA$ for the given sequence $z$ is calculated as:
\begin{equation}
    MSA(z) = [SA_1(z); SA_2(z);\dots;SA_n(z)]W_{msa},
\end{equation}
where $W_{msa}\in\mathbb{R}^{n.K_h\times K}$ denotes the trainable parameters in the multi-head attention network.

The skip connection is applied with this network, outputting sequence representations at different layers ($z_i, i\in {3,6,9,12}$) of the transformer. Their size of $\frac{H\times W\times D}{P^3}\times K$ is reshaped to $\frac{H}{P}\times\frac{W}{P}\times\frac{D}{P}$ tensors. Moreover, a set of $3\times3\times3$ convolutional and normalization layers is applied onto the embedding space tensors to bring them back to the input space for each skip connection.

A deconvolutional layer is used on the output of the last layer of the transformers to increase its shape by a factor of 2. The resized output is concatenated with the output of the preceding transformer layer (e.g. $z_9$) before going into a set of $3\times3\times3$ convolutional layers. This output is then upsampled once again using a deconvolutional layer. All the way up to the top of the network, the process repeats to bring the feature map to the original input size. A $1\times1\times1$ convolutional layer with a softmax activation is then applied on the final representation to output a voxel-wise predictions for segmentation.

\subsection{Loss Function}
The network is supported by a sum of a dice loss as given in Equation~\ref{unetr_dice_loss} and a focal loss as given in Equation~\ref{unetr_focal_loss}. 

\begin{equation}
   \label{unetr_dice_loss}
   \mathcal{L}_{Dice} = \frac{2\sum_{i}^{N} \hat{p_i} y_i}{\sum_{i}^{N} \hat{p_i}^2 + \sum_{i}^{N} y_i^2},
\end{equation}

\begin{equation}
   \label{unetr_focal_loss}
   \mathcal{L}_{Focal} = -\sum_{i}^{N}\alpha y_i (1 - \hat{p_i})^{\gamma}log(\hat{p_i}) - (1 - y_i)\hat{p_i}^{\gamma}log(1-\hat{p_i}),
\end{equation}

\begin{equation}
    \mathcal{L}_{Final} = \mathcal{L}_{Dice} + \mathcal{L}_{Focal}
\end{equation}

where $N$ is the sample size, $\hat{p}$ is the model prediction, $y$ is the ground truth, $\alpha$ is the weightage for the trade-off between precision and recall in the focal loss (empirically set to 1), and $\gamma$ is focusing parameter (empirically set to 2).

\section{Dataset}
\subsection{HECKTOR Dataset}
The HECKTOR challenge~\cite{oreiller2022head,andrearczyk2022overview} provides the dataset of CT, FDG-PET, segmentation masks, tumor bounding box information in a \texttt{csv} file and EHR data. We disregard the EHR data since our task is segmentation using only the imaging data. A total of 325 patient cases (224 training, 101 testing) are provided in the dataset, with the testing set ground truth being inaccessible for competition purposes. The data is multicentric (six centers) and multimodal (CT and PET).

\subsection{Image Preprocessing}
A few preprocessing techniques were used to clean and manage the data. The first was to crop the randomly and massively sized images of CT and PET into $144\times144\times144mm^3$ using the provided bounding box information. Given that this information is provided by the challenge committee, the cropping is accurate, the mapping between modalities is consistent, and the full tumor is within this region. Furthermore, normalization was applied on both CT and PET scans. CT images were clipped within the range of (-1024, 1024) and then normalized to (-1, 1). PET images were normalized using Z-score normalization. Finally, resampling to an isotropic voxel spacing of $1.0mm$ was used for all scans.

\subsection{Data Augmentations}
Several data augmentations were applied on the imaging data to let the model get exposed to variations of the given data. Random rotation in the range of (-45, 45), zooming, elastic deformation, mirroring and gamma correction in the range of (0.5, 2) were experimented with in different combinations. Note that gamma correction is applied only on the PET scans as they are dark in non-tumor regions, and investigating different intensities is supposed to help improve the model. Zooming with a factor of 1.25, reducing the current size down to $115\times115\times155mm^3$, is assumed to accentuate the tumor regions where the tumor is small in size.

\section{Experimental Setup}
For our experiments, we used two NVIDIA RTX A6000 (48GB) GPUs, and the implementation was carried out using PyTorch library. The network has a VIT-B16~\cite{dosovitskiy2020image} as the encoder with $L=12$ layers, a patch resolution of $16\times16\times16$, and $K=768$ embedding size. Two input channels corresponding to CT and PET, and one output channel corresponding to the segmentation mask were used in the network. 

All the experiments with UNETR were trained with a batch size of 8 for 800 epochs. The reason why the model was trained for long was to explore how it would act over long epochs. An AdamW optimizer with a base learning rate of 1e-3 and weight decay of 1e-5, and a cosine annealing schedule, which starts with the base learning rate and reduces it to 1e-5 every 25 epochs, were using in the experiments.

\begin{table}[t!]
    \centering
    \caption{The table shows results of using different data augmentations with our UNETR model. Model performance is presented when using different data augmentations. NA=No Augmentation, MR=Mirroring, RT=Rotation, ZM=Zooming, GC=Gamma Correction, ED=Elastic Deformation.}
    
    \begin{adjustbox}{width=1\columnwidth,center}
    \begin{tabular}{c|c|c|c|c|c|c|c|c}
         & NA & MR,RT & MR,RT,ZM & MR,RT,GC & MR,RT,ED & MR,RT,ZM,GC & MR,RT,ZM,GC,ED & MR,RT,GC,ED \\
        \hline
        DSC & 0.741 & 0.791 & 0.777 & 0.788 & 0.788 & 0.775 & 0.784 & \textbf{0.794} \\ 
        Precision & 0.726 & 0.775 & 0.742 & \textbf{0.778} & 0.768 & 0.767 & 0.765 & 0.761 \\
        Recall & 0.805 & 0.834 & 0.850 & 0.829 & 0.845 & 0.822 & 0.850 & \textbf{0.861}
    \end{tabular}
    \end{adjustbox}
    \label{tab:unetr_augmentations_resultss}
\end{table}

\section{Ablation Studies}
First set of experiments were to explore different combinations of data augmentations that help the model achieve the highest DSC score. Table~\ref{tab:unetr_augmentations_resultss} lists all the combinations that were applied on the data using the model.

Another set of experiments that we did were to implement two leading CNN-based models as in their original work. Squeeze and excitation normalization based residual U-Net proposed by Iantsen et al.~\cite{iantsen2020squeeze} that won the HECKTOR competition in 2020 was selected as the first CNN model, dubbed as SE-based U-Net. An automatically configured U-Net architecture design by Isensee et al.~\cite{isensee2020nnu} that won the BRATS 2020~\cite{menze2014multimodal} was selected as the second model, dubbed as nnU-Net. Both the models were trained and validated as per the corresponding papers using the HECKTOR 2021 dataset.

\section{Results}
\subsection{Quantitative Results}

\paragraph{Augmentation Results.}
The model was trained with several combinations of data augmentations and with no augmentations to verify the essence of such techniques to the transformer model. For this set of experiments, we used the training data, splitting it into 169 cases for training and 55 cases for validation. Table~\ref{tab:unetr_augmentations_resultss} shows the list of combinations and their corresponding results. With no augmentation, the model achieves DSC of 0.741, precision of 0.726, and recall of 0.805. The results improved with all the augmentation combinations in all metrics, proving the essence of augmentations to the model. The set of mirroring, rotation, gamma correction on PET, and elastic deformation yields the highest performance with DSC of 0.794, precision of 0.761, and recall of 0.861.

\begin{table}[t!]
\centering
\caption{Dice, precision and recall of different models on the validation set are shown on the left. The dice, precision and recall of the UNETR model on the testing set are shown on the right. Mean and standard deviation are reported for the three models (left) which were trained and cross validated from scratch to provide a fair comparison.}
\begin{tabular}{ cc }   
    \centering
    
    \begin{minipage}{.7\linewidth}
    \centering
        {\begin{tabular}{l|ccc}
        \multicolumn{4}{c}{\textbf{Validation Set (5-Fold Cross Validation)}} \\
        \hline
         & SE-based U-Net & nnU-Net & UNETR\\
    \hline
        DSC	  & \textbf{0.757}\scriptsize\textpm0.048 & 0.748\scriptsize\textpm0.061 & 0.736\scriptsize\textpm0.043\\
        Precision	  & \textbf{0.748}\scriptsize\textpm0.060 & 0.768\scriptsize\textpm0.095 & 0.766\scriptsize\textpm0.022\\
        Recall  &  0.784\scriptsize\textpm0.044 & \textbf{0.788}\scriptsize\textpm0.031 & 0.766\scriptsize\textpm0.058 \\
        \hline

    \end{tabular}}
    \end{minipage} &
    \begin{minipage}{.22\linewidth}
    \centering
        \begin{tabular}{ c } 
        \textbf{Testing Set} \\
        \hline
        UNETR \\
        \hline
         0.736 \\
        0.773 \\
        0.760 \\
        \hline
        \end{tabular} 
    \end{minipage}
\end{tabular}
\label{tab:unetr_final_results}
\end{table}

\paragraph{Cross Validation Results.}
Once we identified the right set of augmentations (rotation, mirroring, gamma correction, and elastic deformation), we ran a cross validation using leave-one-center-out approach, in which one of the five centers in the training set was used as validation iteratively over all the five centers. The results are summarized in Table~\ref{tab:unetr_final_results}. It achieved DSC of 0.736 (\textpm0.043), precision of 0.766(\textpm0.022), and recall of 0.766(\textpm0.058).

\paragraph{Comparison of Models.}
In comparison to UNETR, we used two CNN based models: SE-based U-Net and nnU-Net. All the three models were trained and validated from scratch using the competition training data in a leave-one-center-out fashion. The results are listed in Table~\ref{tab:unetr_final_results}. UNETR achieved DSC of 0.736 (\textpm0.043) that is only 0.021 and 0.012 lower that the leading CNN models respectively. That verifies that the transformer model is capable of learning and competing closely when trained from scratch. 

\paragraph{Testing Results.}
We are thankful to the organizers of the HECKTOR team for kindly checking the UNETR predictions on the withheld testing set for further verification on the model's capacity. The metrics achieved on the testing set are highly similar to the cross validation results, with DSC of 0.736, precision of 0.773, and recall of 0.760 (compared to 0.736, 0.766, and 0.766 in the validation set respectively).

\subsection{Qualitative Results}
\begin{figure}[t!]
    \centerline{\includegraphics[width=\columnwidth]{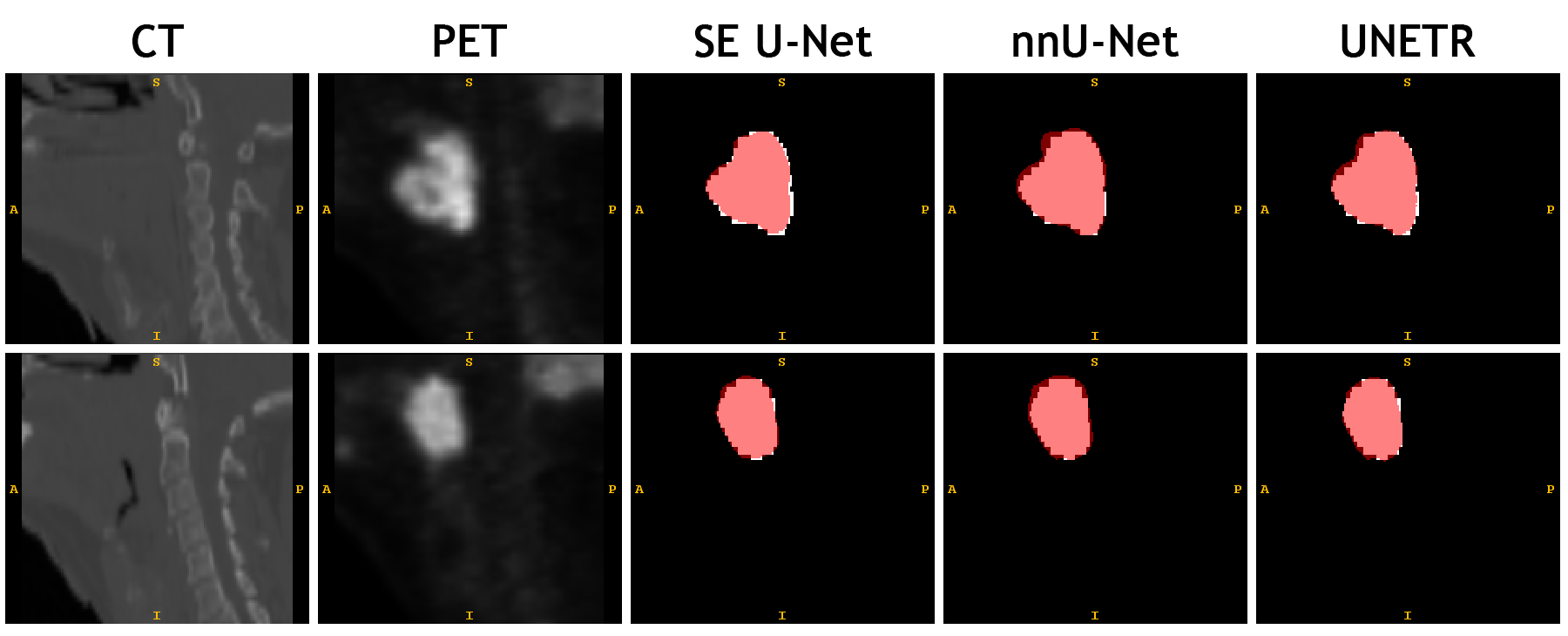}}
    \caption{The figure shows segmentation examples where the models performed well. White represents the ground truth mask and red represents the model's prediction mask. All the models mostly produce these kinds of segmentation results. SE-based U-Net and UNETR segments the tumors very accurately, while nnU-Net over-segments by a small extent.}
    \label{unetr_good_segments}
\end{figure}

\begin{figure}[t!]
    \centerline{\includegraphics[width=\columnwidth]{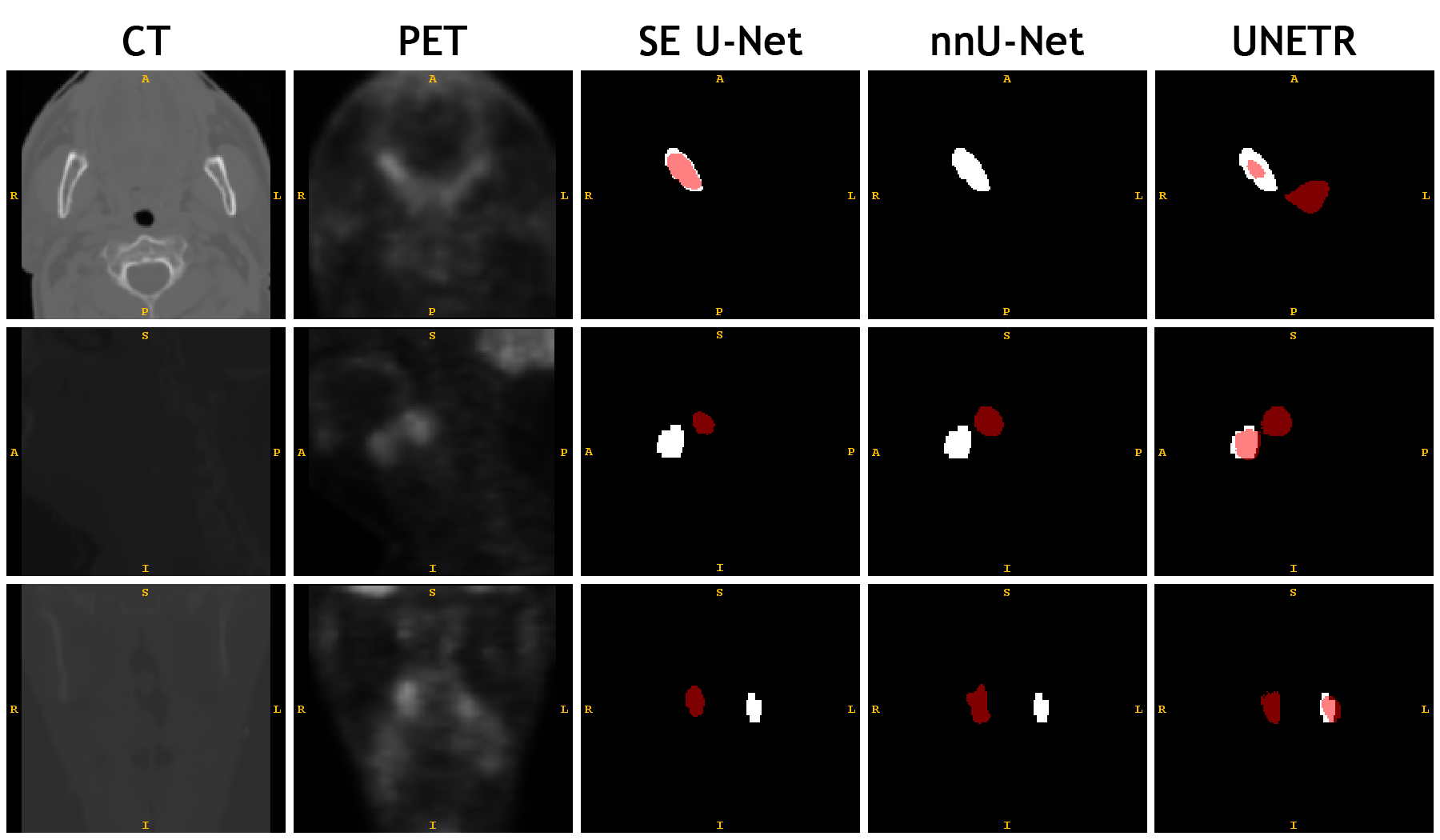}}
    \caption{The figure shows segmentation examples where the models did not perform well. White represents the ground truth mask and red represents the model's prediction mask. The models fail to accurately segment the tumor on account of the unclarity in PET and CT scans and the small size of tumors. UNETR model can partially locate the tumor region in the samples, whereas the other two models fail to do that. SE-based U-Net occasionally shows better output than UNETR.}
    \label{unetr_bad_segments}
\end{figure}

Further comparison and study of the models were conducted via qualitative analysis. The prediction masks from each model were compared against each other and the ground truth to understand how and where the models' outputs differ from each other. Figure~\ref{unetr_good_segments} illustrates CT, PET, ground truth in white and prediction masks from the three models in red. Such a performance of high quality is true in most data samples for all the three models, where the models performed fairly equally well, with only minor differences. Figure~\ref{unetr_bad_segments} depicts cases where the models performed poorly, missegmenting the tumor regions. It is worthy of note that all the three models heavily depend on both the CT and PET scans. CT is responsible for the structural body information that the models need to distinguish different organs, whereas PET provides clarity in intensities of different structures, such as highlighting tumor regions as they are densely structured, as is visible in Figure~\ref{unetr_good_segments}, \ref{unetr_bad_segments}. 

\begin{figure}[t!]
\centering

\begin{subfigure}{0.31\textwidth}
    \includegraphics[width=\textwidth]{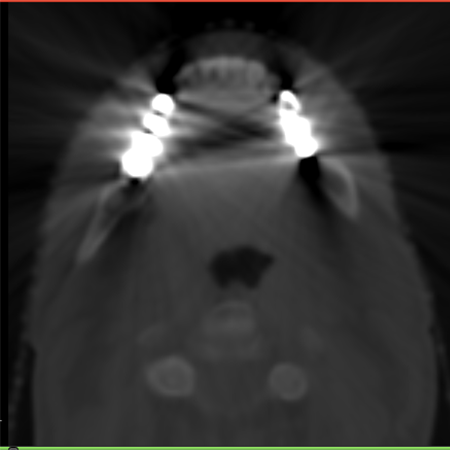}
    \caption{CT}
    \label{fig:unetr_bad_ct}
\end{subfigure}
\hfill
\begin{subfigure}{0.31\textwidth}
    \includegraphics[width=\textwidth]{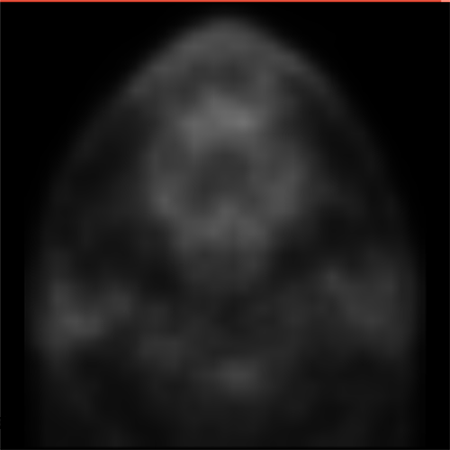}
    \caption{PET}
    \label{fig:unetr_bad_pt}
\end{subfigure}
\hfill
\begin{subfigure}{0.31\textwidth}
    \includegraphics[width=\textwidth]{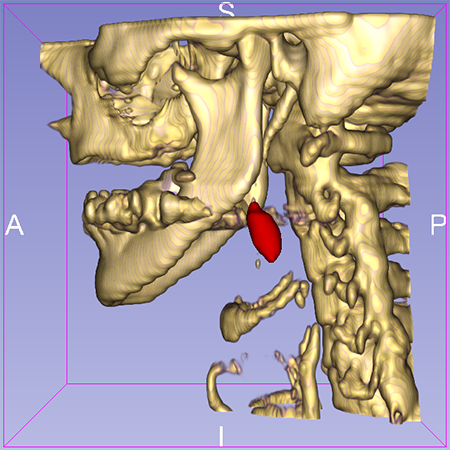}
    \caption{Mask}
    \label{fig:unetr_bad_mask}
\end{subfigure}
        
\caption{The figure depicts one sample with which the models struggled to segment; (\textit{a}) shows a CT slice with artifacts, (\textit{b}) shows an unclear PET slice, and (\textit{c}) shows a small sized mask (in red) superimposed on the CT bone structure. Note that this is a single scan, containing all the three issues.}
\label{fig:unetr_bad_sample_reasons}
\end{figure}

Further study on the poorly segmented cases was conducted to understand the foundational reasons. Figure~\ref{unetr_bad_segments} shows the cases where the models failed to segment accurately. Such issues occur with only a certain group of samples. To understand the struggle of the models, the cases with the lowest DSC scores were extracted and examined. Three major reasons were found as the cause of the failure, as shown in Figure~\ref{fig:unetr_bad_sample_reasons}. First, as depicted in Figure~\ref{fig:unetr_bad_ct}, most of the poorly segmented scans suffer from streak artifacts in CT scans introducing irregularities in data, thereby causing the model confusion. Such artifacts are assumed to be caused by dental implants as they occur in the tooth area. The second reason is that the intensity values for different structures in PET scans in these cases are not well-outlines, especially in the tumor regions. This causes the models to produce faulty output, segmenting other regions with higher intensities. Such a sample can be seen in Figure~\ref{fig:unetr_bad_pt}. Lastly, the tumor sizes in such scans are relatively small, especially when compared to well-performing cases, as depicted in Figure~\ref{unetr_bad_segments},~\ref{fig:unetr_bad_mask}. The models struggle to figure out where to focus on when they encounter the detection of tiny regions with a largely abundant data in the background. 

\section{Discussion}
To further discuss on the results of the data augmentations applied for the UNETR model, we can say that all of them were fruitful for the model's improvement. The combination with the zooming with a factor of 1.25 did not show much potential. It is hypothesized that zooming would help preserve information and highlight the tumor regions, especially the ones with small sizes, however, it did not show much effectiveness with the transformer backbone. 

The comparison of the models shows that the CNN-based models outperform the ViT-based counterpart by a slight extent. This, however, should be an indicator that, even with the limited data, the transformer model can perform well, achieving promising results. Its consistency on the withheld testing set, while verifying the cross-validation, manifests that the model is indeed learning properly (i.e. no overfitting, no data leakage, etc.) even when trained from scratch.

Qualitative analysis demonstrates that the three models are side-by-side in terms of the performance, with slight differences in over- or under-segmentation. It also shows that there are certain reasons as to why the models struggled to perform the segmentation, including the artifacts in CT, obscureness in PET, and small tumor sizes.

\section{Conclusion}
H\&N cancer is one of the most common types of cancer, and automatic segmentation of H\&N tumor is an essential task that should be investigated to a greater extent. This work focused on the transformer-based model for the task for the first time, and compared its performance against two powerful CNN models. We showed that the model performs comparable to the CNNs, achieving a mean DSC of 0.736 (\textpm0.043), a precision of 0.766(\textpm0.022), and a recall of 0.766(\textpm0.058) when trained from scratch. On the HECKTOR testing set, the model yielded similar results, with DSC of 0.736, precision of 0.773, and recall of 0.760. 

Reported results of the transformer network are slightly lower than the well-mature CNN counterparts (0.021 and 0.012 for SE-based U-Net and nnU-Net respectively). Although this is the case, we believe that CNNs have undergone a number of improvements in structure, design, subcomponents, etc., whereas the transformers are yet to be studied in more detail. This chapter is one of the early works to compare the two designs, different in nature, for the segmentation task. Furthermore, pretraining via self-supervised learning has been noted numerous times as the key reason for the success of transformers in the NLP domain, and therefore, we believe that the self-supervised approach to the task should be studied, particularly in the case of transformer-based models with limited data for image segmentation. 


\chapter{Segmentation with Super Images: A New 2D Perspective on 3D Medical Image Analysis}
\label{chp_superimages}


\section{Introduction}
The latest advances in DL have shown their efficacy in numerous computer vision tasks such as classification~\cite{classification}, detection~\cite{detection,detection2} and segmentation~\cite{segmentation,segmentation2,segmentation3}. Medical imaging analysis was no exception in availing itself with the benefits of the CNNs. The availability of medical datasets helped the field progress even faster. Automation of many medical tasks has been attracting researchers to explore the field to deeper extent.

Numerous DL solutions have been proposed for different tasks within the medical sphere. In the segmentation task, U-Net~\cite{ronneberger2015u} is considered the most widely utilized model, which was later augmented with many additional features~\cite{cciccek20163d,feng,iantsen2020squeeze,zhang2018road}. ViT~\cite{dosovitskiy2020image} introduced a completely new way to look into such problems, with its many applications~\cite{chen2021transunet,zhang2021transfuse,transbts} for segmentation.

The nature of medical images is different from natural counterparts, being volumetric in many modalities such as CT, PET and MRI. This factor contributed to a still-ongoing discussion over the design of architectures for medical images. Should we enforce using 2D designs for three dimensional data? Or, should we leave this idea, and move to more complex 3D designs? Or, is there in-between consensus?

Those who support designing 3D networks for volumetric data claim that the depth information is crucial for model learning and can be captured only by 3D networks~\cite{3dcompact,cciccek20163d,3dvnet}. 2D counterparts lack this attribute, and therefore, cannot perform as well as the 3D models. Another argument they put forth is that the nature of 3D data is similar to real life, which is why 3D models achieve better performance~\cite{3dcompact,3dvnet}. However, the drawback to using such networks is that 3D convolutional, max pooling and up-convolutional operations are performed instead of 2D which are by design more complex, requiring more computational power and training/inference time~\cite{2dguy,super}.

Those who support designing 2D networks argue that such models are much more cost- and time-effective as well as offers more options to apply transfer learning~\cite{super,2dreinvent,2dnode}. Pretraining models on large-scale datasets like ImageNet~\cite{imagenet} and then fine-tuning it with the dataset at hand can be considerably beneficial to model learning, especially when the medical dataset is small in quantity. Another benefit to supporting 2D networks is that volumetric data can be converted to 2D slices, generating a larger quantity of data. Finally, there are more 2D architectural options in choosing an encoder for U-Net-like models, which makes it easier to customize and adjust new models to the need of the problem at hand~\cite{2dguy}.

Certainly, there is a group who claim that the hybrid of 2D and 3D in various fashions can bring out the best of the two approaches~\cite{25conv,2dnode,25conv2}. However, in reality it proves not to be the case since hybrid approaches are fundamentally failing to capture innate benefits of 2D and 3D architectures~\cite{2dreinvent}.

This chapter introduces a new perspective to dealing with three dimensional data in a 2D fashion. We generate two dimensional \textbf{super images (SIs)} from a 3D input by stacking the depth (i.e. slices) side by side, and we train a 2D network for the segmentation task. A similar work was proposed in~\cite{super} on natural videos; however unlike them, this work newly introduces the idea to the medical field and volumetric data in particular. This novel approach can produce comparable results to 3D networks and we believe it lays the foundation for a new look into using volumetric medical data. The main contributions of the chapter are:
\begin{itemize}
    \item We introduce a new perspective to using biomedical three dimensional data by casting them into super images and training 2D networks with them;
    \item We empirically show that pretraining and preprocessing techniques on top of using super images can easily improve the model performance;
    \item We validate the proposed method on different datasets of CT, PET, and MRI to show the viability of the approach.
\end{itemize}

\section{Methodology}
\subsection{Datasets and Preprocessing}
To validate the new method, we applied it on two different datasets: HECKTOR 2021 challenge~\cite{andrearczyk2022overview} and atrial segmentation challenge~\cite{asc} datasets. The datasets are different in modality, organ-at-risk, and dimensions; therefore, they are preprocessed accordingly.

\subsubsection{Head and Neck Tumor}
\begin{figure}\centering
\begin{tabular}{ccc}
{\includegraphics[width=0.3\columnwidth]{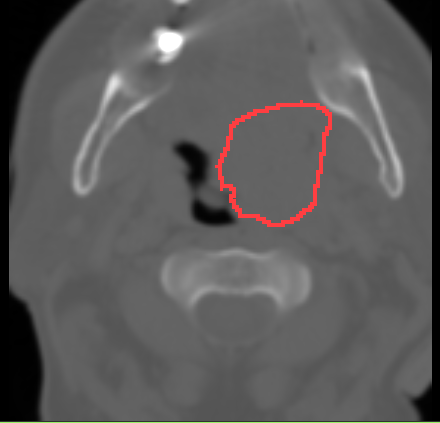}}&
{\includegraphics[width=0.3\columnwidth]{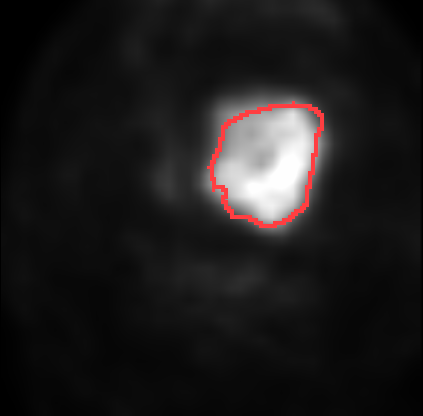}}&
{\includegraphics[width=0.3\columnwidth]{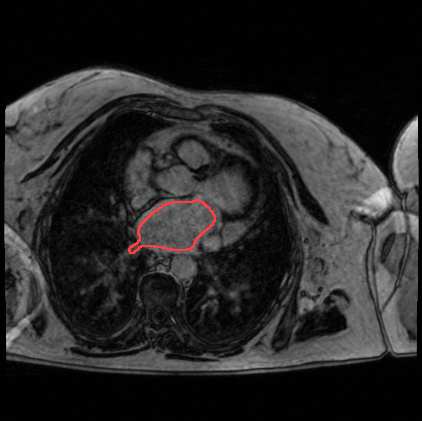}}
\\
(a)&(b)&(c)
\end{tabular}
\caption{The figure shows sample slices from the two datasets. (a) and (b) depict CT and PET scans with red region highlighting tumor from the HECKTOR dataset respectively; and (c) shows an MRI slice with the red region corresponding to atrium from the Atria segmentation dataset.}
\label{sample}
\end{figure}

HECKTOR dataset\footnote[1]{aicrowd.com/challenges/miccai-2021-hecktor} comprises 224 training CT and PET scans of patients diagnosed with head and neck tumor. Bounding box information is also available for tumor localization, which was used to crop the scans and masks down to the size of $144\times144\times144mm^3$ with consistency between the images. Sample CT and PET slices are depicted in Figure~\ref{sample} (a) and (b) respectively. Here, red lining correspond to the tumor region. The images were further cropped down to the size of $80\times80\times48$ for a faster performance as ablation studies. This clearly highlights the tumor region and helps the model learn more effeciently. Additionally, both the CT and PET scans were resampled to have an isotropic voxel spacing of $1mm^3$ and their intensity values were normalized. The CT scans were clipped within $(-1024, 1024)$ and normalized to $(-1, 1)$; and the PET scans were normalized using \textit{Z}-score normalization.

\subsubsection{Atrium}
Atrial segmentation challenge dataset\footnote[2]{atriaseg2018.cardiacatlas.org/data} comprises 100 gadolinium contrast (GE) MRIs in the training set. Figure~\ref{sample} (c) shows a sample slice of a scan with red lining corresponding to the atrial region. The dimensions of the scans are differing from patient to patient, and they were all resized to the same size of $512\times512\times82mm^3$. Intensity normalization was also applied to the images. No further data augmentation was applied on the dataset unless reported otherwise. 

To show the applicability and generalizability of the proposed idea, we intentionally chose two datasets with different modalities, with CT, PET and MRI. On top of that, the tissues that are to be segmented are different; HECKTOR dataset is tumor segmentation and the second dataset is atrial segmentation. This shows the idea is not limited to a specific tissue type.

\section{Super Image Generation}

\begin{figure}[htbp]
\centerline{\includegraphics[width=\columnwidth]{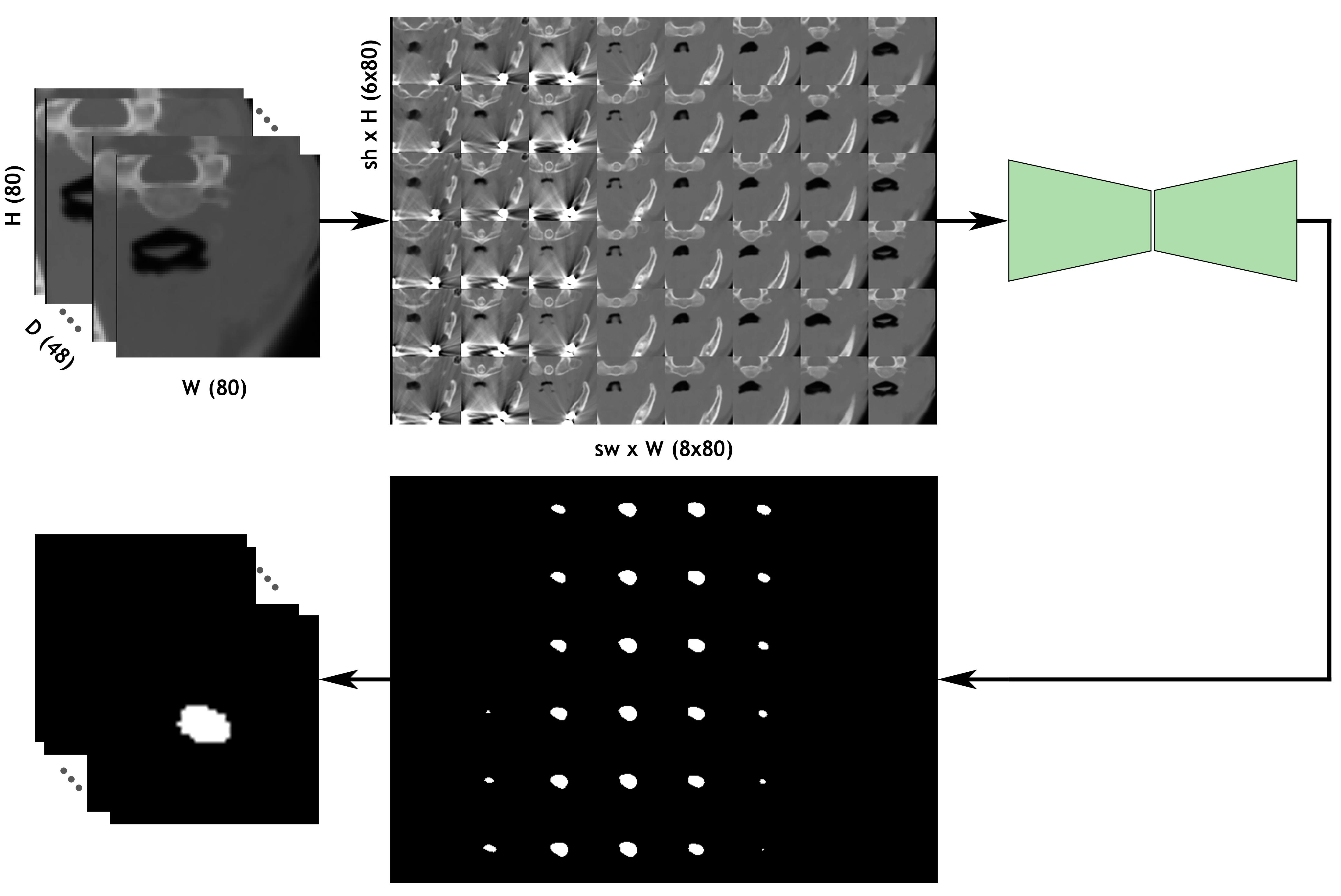}}
\caption{The figure shows the construction of super images from volumetric data. We rearrange the depth dimension by assembling the slices together to generate the super image. It is then fed to a 2D segmentation network. The model yields the prediction mask which is then rearranged back to the original shape. Note that the volumetric prediction mask shows a tumor region for visualization purposes.}
\label{super_image}
\end{figure}

A 3D network is claimed to capture and learn from features extracted from the depth information since they use three dimensional kernels; however, we assume that these features are still detectable and learnable in two dimensional SIs by well-designed deep neural networks. To that end, we generate SIs from volumetric data by slicing and stacking these slices side by side in order, as depicted in Figure~\ref{super_image}.

Given a 3D image $x\in \mathbb{R}^{H\times W\times D\times C}$, where $H$ is the height, $W$ is the width, $D$ is the depth, and $C$ is the number of channels, the depth dimension is rearranged. The resulting image $s\in \mathbb{R}^{\hat{H}\times\hat{W}\times C}$ is now 2D, where $\hat{H} = H\times sh$, and $\hat{W} = W\times sw$; $sh$ and $sw$ represent the degree by which the height and width should be rearranged respectively to generate a grid size of $sh\times sw$. As a demonstration, the size of $80\times80\times48\times2$ (2 for both CT and PET slices), having $48$ as the depth, can be considered with $sh$ of 6 and $sw$ of 8, thus generating the SI in the dimensions of $480\times640\times2$, as shown in Figure~\ref{super_image}. 2D U-Net (or any other 2D segmentation model for that matter) can then be applied on these SIs to perform the segmentation. The model output in the SI form is finally cast back to the original size of the input scan.

\section{Implementation Details}
Two NVIDIA RTX A6000 GPUs were used to perform the experiments, and PyTorch library was used for implementation. We ran all the experiments for 100 epochs. We used AdamW optimizer with a base learning rate of 1e-3 and weight decay of 1e-5, and cosine annealing scheduler that starts with the base learning rate, lowering it to 1e-5 in 25 epochs before setting it back to the base learning rate. The batch size was set to 4, 8, or 16 depending on the dataset and architecture. 

\section{Experiments and Results}

\begin{table}\centering
\caption{The table shows the results of vanilla 3D U-Net (comparison target) to SI-based 2D U-Net on the HECKTOR training/validation dataset. The results are the mean and standard deviation of 5-fold cross validation.}\label{hecktor_results}
\setlength{\tabcolsep}{3pt}
\begin{tabular}{|l|l|l|l|l|l|l|c|c|}
\hline
Model &  Image Size & \textit{sh} & \textit{sw} & DSC & Precision & Recall & Params & MACs(G) \\
\hline
3D & $144\times144\times144$ & - & - & 0.718\scriptsize\textpm 0.055 & 0.749\scriptsize\textpm 0.050 & 0.747\scriptsize\textpm 0.0675 & 3.61M & 518.61 \\
2D & $144\times144\times144$ & 12 & 12 & 0.700\scriptsize\textpm 0.070 & 0.731\scriptsize\textpm 0.046 & 0.731\scriptsize\textpm 0.075 & 1.21M & 319.98  \\
\hline
3D & $80\times80\times48$ & - & - & \textbf{0.779}\scriptsize\textpm 0.031 & 0.787\scriptsize\textpm 0.021 & 0.822\scriptsize\textpm 0.039 
& 3.61M & 53.35 \\
2D & $80\times80\times48$ & 8 & 6 & \textbf{0.778}\scriptsize\textpm 0.033 & 0.799\scriptsize\textpm 0.021 & 0.810\scriptsize\textpm 0.044  & 1.21M & 32.92 \\
2D & $80\times80\times48$ & 6 & 8 & 0.777\scriptsize\textpm 0.034 & 0.793\scriptsize\textpm 0.018 & 0.816\scriptsize\textpm 0.044 & 1.21M & 32.92\\
2D & $80\times80\times48$ & 12 & 4 & 0.770\scriptsize\textpm 0.030 & 0.809\scriptsize\textpm 0.037 & 0.801\scriptsize\textpm 0.055 & 1.21M & 32.92\\
2D & $80\times80\times48$ & 4 & 12 & 0.759\scriptsize\textpm 0.043 & 0.790\scriptsize\textpm 0.016 & 0.797\scriptsize\textpm 0.062 & 1.21M & 32.92 \\
2D & $80\times80\times48$ & 24 & 2 & 0.744\scriptsize\textpm 0.044 & 0.765\scriptsize\textpm 0.027 & 0.809\scriptsize\textpm 0.047 & 1.21M & 32.92 \\
2D & $80\times80\times48$ & 2 & 24 & 0.762\scriptsize\textpm 0.035 & 0.779\scriptsize\textpm 0.023 & 0.809\scriptsize\textpm 0.052 & 1.21M & 32.92 \\
\hline
\end{tabular}
\end{table}

We applied the idea on two datasets. The ground truth for the testing set of both datasets were unavailable for competition purposes; therefore, \textit{k}-fold cross validation on the training set was used for all experiments.

The HECKTOR dataset was explored in two settings: (a) with the initial size of $144\times144\times144mm^3$, and (b) the cropped size of $80\times80\times48mm^3$. In both settings, 5-fold cross validation was used, and the mean values along with standard deviations are reported in Table~\ref{hecktor_results}. 

In the first setting, we trained and compared two models from scratch: 3D U-Net with volumetric data and 2D U-Net with SIs. DSC of 0.718, precision of 0.749, and recall of 0.747 were achieved with 3D U-Net. The 2D model using SIs achieved marginally lower results than its 3D counterpart, with DSC of 0.700, precision of 0.731 and recall of 0.731. 

In the second setting, the 3D model was trained as a target for the SI-based 2D model. The 3D network reached the mean DSC of 0.779, precision of 0.787, and recall of 0.822. For the 2D network, differently sized SIs were used; i.e. different values were used for $sh$ and $sw$ as listed in Table~\ref{hecktor_results}. We generated SIs with $sh$ and $sw$ set to $8\times6$, $6\times8$, $12\times4$, $4\times12$, $24\times2$, and $2\times24$ grid sizes before feeding into 2D U-Net. Note that the grid sizes are the multiples of the depth dimension since this dimension is rearranged to create SIs. Table~\ref{hecktor_results} shows that the SI-based network with the most square-like shape, i.e. grid sizes of $8\times6$ and $6\times8$, produced the highest metric values, with DSC of 0.778 and 0.777 respectively. This is highly comparable to the 3D model. The performance for SIs with more disproportionate ratios was slightly worse, with DSC range of 0.744 to 0.770. This set of experiments prove that the generation of SIs with similar $sh$ and $sw$ is more favorable for better performance.

\begin{table}\centering
\caption{The table shows the results of vanilla 3D U-Net (comparison target) to SI-based 2D U-Net on the atrial segmentation training/validation dataset. The results are the mean and standard deviation of 4-fold cross validation. PT stands for 2D U-Net pretrained on ImageNet1k, and A stands for augmentations.}\label{asc_results}
\setlength{\tabcolsep}{6.3pt}
\begin{tabular}{|l|l|l|l|l|l|l|}
\hline
Model &  Image Size & \textit{sh} & \textit{sw} & DSC & Precision & Recall \\
\hline
3D U-Net & $512\times512\times88$ & - & - & \textbf{0.893}\scriptsize\textpm 0.011 & 0.898\scriptsize\textpm 0.011 & 0.894\scriptsize\textpm 0.024 \\
2D U-Net & $512\times512\times88$ & 11 & 8 & 0.812\scriptsize\textpm 0.047 & 0.902\scriptsize\textpm 0.038 & 0.785\scriptsize\textpm 0.050 \\
\hline
2D U-Net & $512\times512\times64$ & 8 & 8 & 0.851\scriptsize\textpm 0.039 & 0.913\scriptsize\textpm 0.018 & 0.822\scriptsize\textpm 0.063 \\
PT & $512\times512\times64$ & 8 & 8 & \textbf{0.895}\scriptsize\textpm 0.013 & 0.872\scriptsize\textpm 0.092 & 0.878\scriptsize\textpm 0.035 \\
PT\&A & $512\times512\times64$ & 8 & 8 & \textbf{0.901}\scriptsize\textpm 0.008 & 0.919\scriptsize\textpm 0.018 & 0.890\scriptsize\textpm 0.029 \\

\hline
\end{tabular}
\end{table}

The atrial segmentation dataset contains only 100 MR images; therefore, 4-fold cross validation is used, leaving more data to the validation set for more strict training. We understood with the HECKTOR dataset that using similar $sh$ and $sw$ is better for performance. With the atrial dataset, we first perform the vanilla U-Net comparison of 2D and 3D, and secondly, we apply simple preprocessing techniques to see how far the model can reach. The first setting simply shows the generalizability of the idea in a new tissue type, whereas the second setting shows how image preprocessing techniques can easily be employed and can boost the vanilla network performance. 

Vanilla U-Net in 2D and 3D comparison was carried out using the MR scans with the dimensions of $512\times512\times88$. Table~\ref{asc_results} shows that 3D U-Net yielded DSC of 0.893, precision of 0.898, and recall of 0.894, whereas the 2D network performed much more poorly (DSC of 0.812) with the grid layout of $11\times8$. This is, as hypothesized, due to the disproportionate layout of the SIs. As such, we dropped a few slices from either end of the depth (i.e. cropped off the same number of slices from both sides of a scan). Specifically, the images were clipped to have 64 slices. This helps with the volume reduction for faster training/inference time, and more importantly, with the square-like formation of the SIs. With that, the model's performance jumped to 0.851 in DSC.

Further techniques included applying pretraining weights and augmentations to see how far we can stretch the performance. The 2D model was initialized with ImageNet1k pretrained weights (shown as PT in Table~\ref{asc_results}) instead of randomly initializing it. Even in the medical domain, weights pretrained on natural images showed a significant improvement with DSC reaching 0.895, overshadowing the 3D model. Lastly, a set of augmentations were applied on top of the pretraining technique. Random flip, random affine, random elastic deformation, random anisotropy, and random gamma were all used as a heavy set of augmentations. Note that these augmentations are only specific for this experiment. These aggressive augmentations pushed all the three metrics, with DSC of 0.901 compared to the baseline DSC of 0.812. The small and easy sets of techniques such as these prove to be highly useful for the model's performance. 

The two models were compared in terms of the number of learnable parameters and multiply–accumulate operations (MACs). The vanilla 3D U-Net has 3.61M parameters, which is three times the number of parameters of a 2D model with only 1.21M. In terms of operations, the 3D model calculates to have 518.61 GMACs as compared to 319.98 GMACs for the 2D counterpart when the HECKTOR dataset with the initial size of $144\times144\times144$ is considered. The values naturally decrease with smaller size setting as shown in Table~\ref{hecktor_results}.

\section{Qualitative Analysis}
\begin{figure}\centering
\begin{tabular}{ccc}
{\includegraphics[width=0.4\columnwidth]{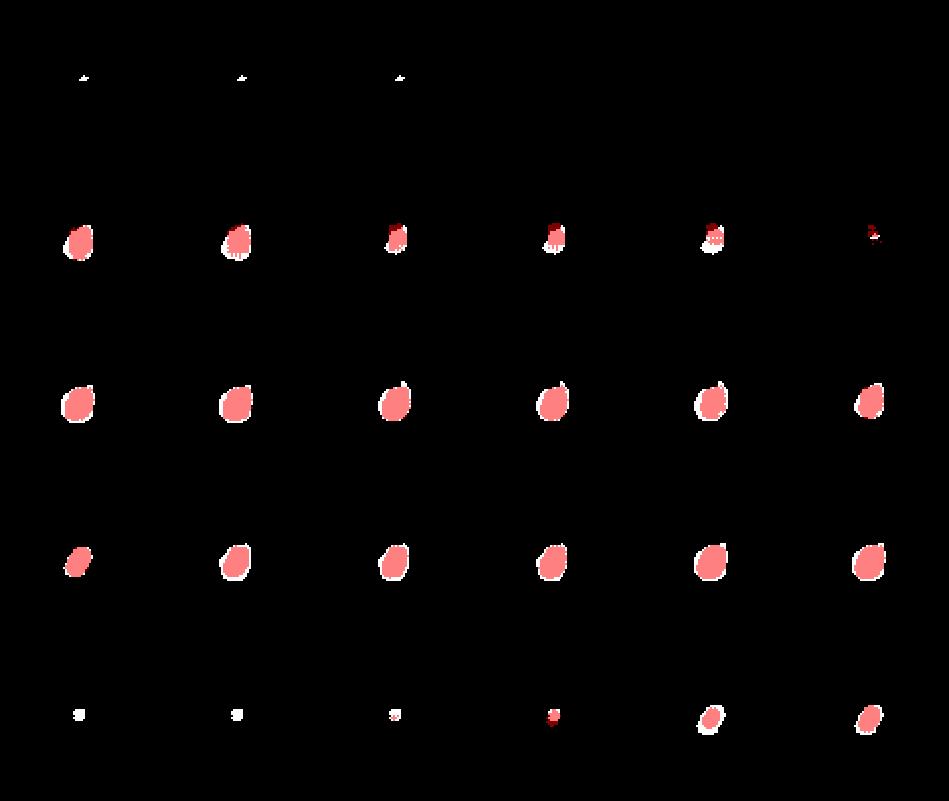}}&
{\includegraphics[width=0.4\columnwidth]{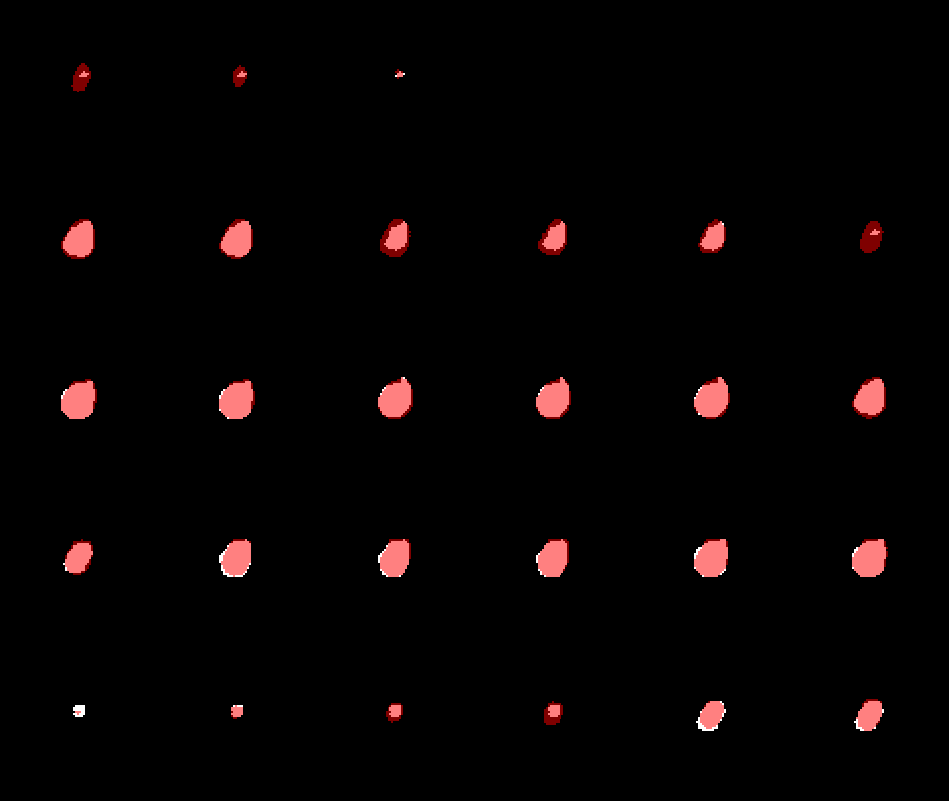}}
\\
(a)&(b)
\end{tabular}
\caption{The figure shows qualitative results on 3D U-Net (on volume) and 2D U-Net (on SI) segmentation results on HECKTOR dataset sample respectively. White is the ground truth and red represents the prediction mask. Note that 3D U-Net results were cast to an SI form after its prediction for full-view comparison.}
\label{qual}
\end{figure}

We conducted qualitative analysis to gain further insight into how 3D model on volume data and 2D model on SIs are differently performing. Figure~\ref{qual} illustrates the segmentation results for 3D (Figure a) and 2D (Figure b) models on the HECKTOR dataset sample respectively. The sample we chose here is the one both models perform objectively well. The white region on the figures represent the ground truth, and the red region represents model predictions. Note that the output from the 3D network was cast to SI form to compare it with its counterpart in a full-view.

As is shown, both models output very similar results for this sample. 2D U-Net oversegments the slices with tumor edges, whereas the 3D model undersegments or completely misses them. The central areas of the tumor is well segmented by the 3D network while 2D again oversegments, causing for a drop in metrics of performance. This phenomenon is common across the dataset and is believed to be caused by the non-volumetric characteristic of SIs.

\section{Discussion}
The proposed idea of converting a volumetric problem into a 2D form was validated through two different types of datasets. At a glance, we see that 2D U-Net using this approach can yield results comparable to 3D U-Net. In the head and neck tumor segmentation problem, the 2D model with SIs produced results slightly lower than the 3D model when the original size of $144\times144\times144$ was used. It is because the task in and of itself is very challenging. With its high dimensions, the tumor regions appear very small as compared to the background. Additionally, when we visually analyzed the dataset, we found several scans with artefacts in the teeth area; and these are the scans the both models are struggling with commonly. Such an artefact is depicted in Figure~\ref{super_image}. 

On the upside, the 2D model performs well with the edges of tumor even if they look tiny in the formation of SIs, as is seen in Figure~\ref{qual}. Although counter intuitive, the 3D model misses such tiny regions for its strict predicting capability, and as it receives the volumetric data in a cubic form, it is assumed that it is unable to accurately pinpoint the contouring edges. It is for its strict predicting feature that its overall performance is slightly higher than its 2D counterpart.

The atrial dataset, with MR images and atrium as organ-of-target, is an entirely different segmentation task from the HECKTOR one. Yet, it did not pose a challenge to both models. Employing basic preprocessing and pretraining techniques, the model's performance using SIs improves, with around 9 percent increase in DSC. Such techniques as applying pretrained weights or clipping the scan to have a square-like form for SIs can help the model drastically improve.

Methodically walking through the problem, we present four arguments as to why it might be favorable to use SIs and 2D networks over using 3D counterparts. First, pretrained weights on large-scale natural images for 2D models are more widespread and easily implementable, both for the encoder and the encoder-decoder models. The performance boost using ImageNet1k weights is illustrated in our experiments too. Pretraining the model on large-scale natural datasets and finetuning on medical applications is much easier with SIs. Second, 2D approach in general offers more options for easily employable data augmentations. Although not explored in details in this work, 3D augmentations are either not yet available or too costly to implement~\cite{santhakumar2021exploring}. It is widely known, and is partially shown in our work, that picking the right set of augmentations is imperative to model learning. Third, a higher number of 2D networks are generally available because of natural images and larger computer vision community. Such networks initially are introduced in the 2D world before being employed in medical imaging tasks. Implementation of these models becomes much easier with SIs when working with volumetric data. Fourth, it is much simpler and quicker to implement SSL on 2D data and 2D networks compared to 3D. Additionally, there is a higher availability of SSL pretrained models in the 2D community; thus studying this aspect of SIs is believed to be a crucial next step.

\section{Conclusion}
When we work on the segmentation task in medical imaging, we often face volumetric data. 3D networks such as 3D U-Net and its variations have been explored intensively for this task; and their performances are indeed impressive. However, the downside to that is they are generally computationally expensive and time-consuming. In this chapter, we studied a new approach of working with volumetric data by casting them to super images and using 2D networks for similar performance. In the HECKTOR dataset, a vanilla 2D U-Net with SIs achieved results comparable to powerful 3D U-Net results. Similarly, in atrial segmentation dataset, the idea is validated with promising performance and potential, especially when the approach is enhanced with various preprocessing techniques. We believe that there is potential for the new perspective of SIs when working with 3D medical data. We will look into further preprocessing techniques, stronger 2D networks such as ViT or deeper CNNs, and self-supervised learning as next steps to improve overall performance of using super images.

\newcommand{\package}[1]{\textbf{#1}} 
\newcommand{\cmmd}[1]{\textbackslash\texttt{#1}} 

\chapter{Tumor Segmentation and Survival Prediction of Patients with Head and Neck Cancer}
\label{chp_prognosischapters}

The prognosis of H\&N cancer patients using DL approaches is understudied. The task by its very nature is very challenging. The previous study results are not encouraging and impractical in clinical use. In this chapter, we propose two distinct solutions~\footnote[3]{This work has a filed patent in the US (USPTO: 17849943)} for the prognosis task. 

\begin{figure}[htbp]
    \centering
    \includegraphics[width=1\textwidth]{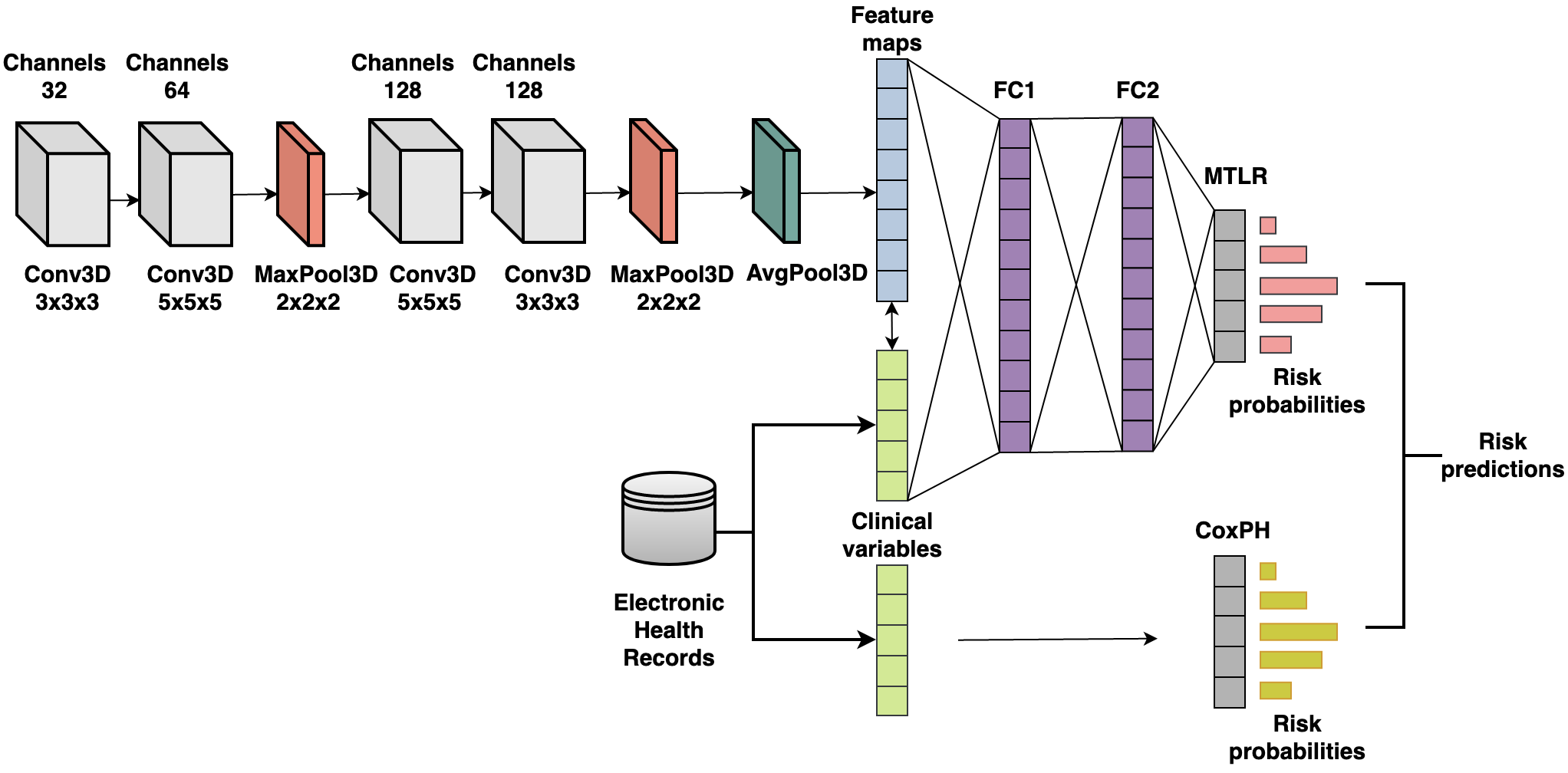}
    \caption{Overall architecture of Deep Fusion V2. Features are extracted from fused CT and PET scans using the CNN network and concatenated with the EHR features. Then, the output is passed to the FC layers before MTLR. Lastly, risk scores from MTLR and CoxPH models are averaged to get the final risk predictions.}
    \label{fig:network_fusion}
\end{figure}

The first is an ensemble network of MTLR, CoxPH and CNN models, dubbed as \textbf{Deep Fusion}, to prognose patients with H\&N cancer using their clinical and imaging (CT and PET) data. The network architecture for Deep Fusion is depicted in Figure~\ref{fig:network_fusion}. We use CNNs to extract features from the CT and PET, and fuse them with the clinical data to make the risk predictions. We trained and tested the proposed solution on 224 and 101 patient cases respectively. The model is tested on the HECKTOR 2021 testing set and achieved the C-index of 0.72, winning the competition.    

\begin{figure}[htbp]\centering
{\includegraphics[width=\columnwidth]{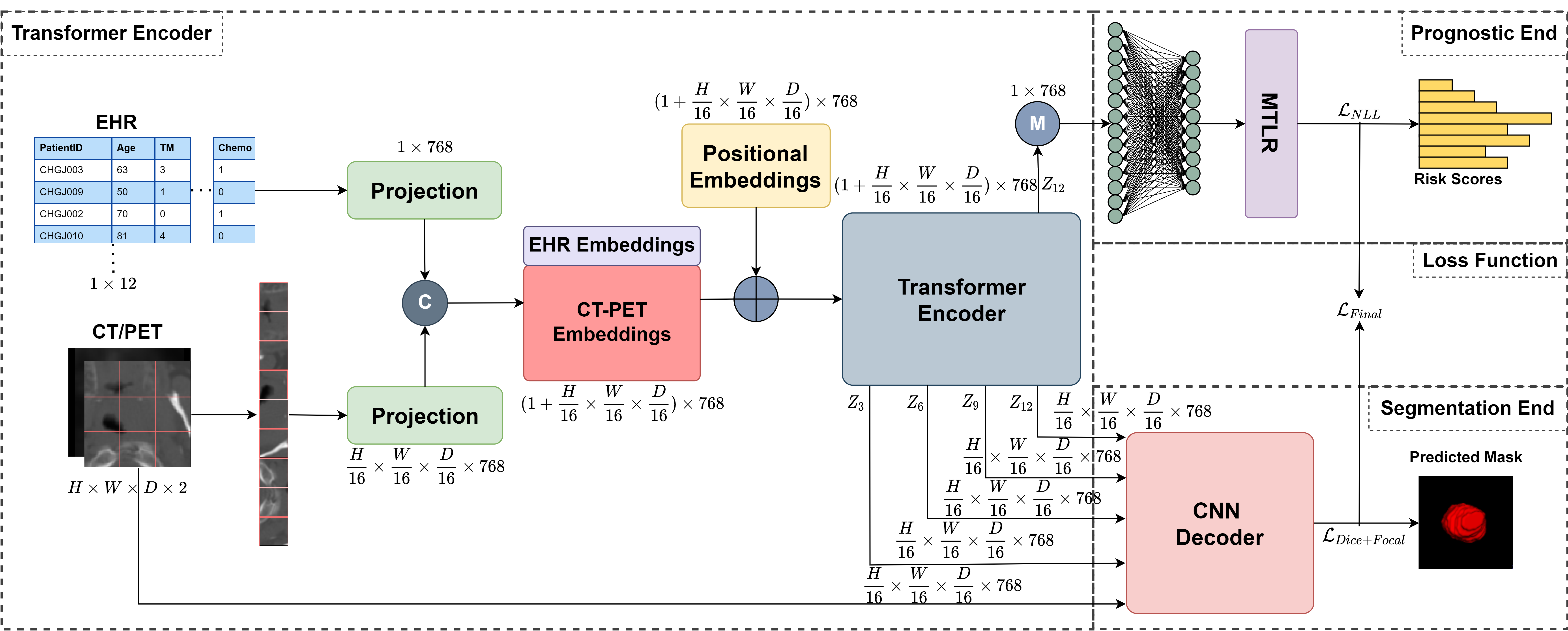}}
\caption{An illustration of the proposed TMSS architecture and the multimodal training strategy. TMSS linearly projects EHR and multimodal images into a feature vector and feeds it into a Transformer encoder. The CNN decoder is fed with the input images, skip connection outputs at different layers, and the final layer output to perform the segmentation, whereas the prognostic end utilizes the output of the last layer of the encoder to predict the risk score.}
\label{tmmsarchitecture}
\end{figure}

The second solution is an end-to-end \textbf{T}ransformer based \textbf{M}ultimodal network for \textbf{S}egmentation and \textbf{S}urvival (\textbf{TMSS}) prediction, presented in Figure~\ref{tmmsarchitecture}. This architecture benefits from the use of transformers for the fact that they can handle different data modalities. The model is designed to perform both the segmentation and prognosis tasks. The same data is used in a five-fold cross-validation form, outperforming the previous prognosis models with C-index of 0.763 and achieving DSC of 0.772 that is comparable to a standalone segmentation model.




\chapter{Thesis Conclusion}
\label{chp_conclusion}
\vspace{-10mm}

Cancer is one of the most lethal diseases in the world, and H\&N cancer is considered a severe type of cancer with a significant number of deaths. While doctors spend hours to manually segment the tumor region from medical imaging data and then perform prognosis using that outcome and other clinical data, traditional ML and DL approaches can automate both these tasks, yielding comparable results instantaneously. This thesis work focuses on providing different solutions to the problem. The segmentation task was studied in depth in two ways: exploring the use of leading CNNs and ViTs for the task, and proposing the concept of super images with 2D networks for segmentation with such volumetric data. Patient outcome prediction task was studied using an ensemble of traditional ML and CNN networks to set the state-of-the-art on the task. A ViT-based model proposed later renews this state-of-the-art because of its capability to handle multimodal data and attend to the right modality while learning. 

The work is highly detailed from different aspects; however, there still exist certain limitations. The primary limitation is that the work is predominantly focused on head and neck cancer, using only the HECKTOR dataset. Further verification of the proposed ideas, especially the super images concept, on external datasets is essential. Generalizability of the concept on other various tasks, such as classification and detection should be tested. Another strong limitation of the whole work is the clinical verification for all the models. Although the prognosis and diagnosis models from Chapter~\ref{chp_prognosischapters} are under a patent, they were not yet tested on a clinical trial. This should be an imperative component of the work since we are trying to propose to solve the issue at a level of clinician's precision. 

For future work, Chapter~\ref{chp_transformers} can benefit from exploring more recent ViT models such as Swin UNETR for faster and more desirable results. More ablation studies could strengthen the argument of using ViTs for such problems. Chapter~\ref{chp_superimages}, as mentioned, should be enhanced with more experiments using deeper CNN and transformer models to improve its results. Although seemingly reaching comparable results to its 3D counterparts, the study of SIs in more depth is believed to be a promising field of research, especially when it is supported by SSL pretraining. Chapter~\ref{chp_prognosischapters} survival analysis can further be boosted with more exploration into the multimodal data fusion approaches so that it becomes favorable even in clinical practice. While intuitive, ViT data embedding and positional encoding can further be analyzed in various fashions to understand the efficacy of the modules within the network. Finally, more recent encoders can be explored to reach even higher performance.

\addcontentsline{toc}{chapter}{\textbf{Nomenclature}}

\renewcommand{\nomname}{Nomenclature}
\renewcommand{\nomAname}{\textbf{\large Abbreviations}}
\renewcommand{\nomGname}{\textbf{\large Mathematical Symbols}}
\renewcommand{\nomXname}{\textbf{\large Superscripts}}
\renewcommand{\nomZname}{\textbf{\large Subscripts}}

\printnomenclature
\cleardoublepage
\phantomsection 


\renewcommand*{\bibname}{References}

\addcontentsline{toc}{chapter}{\textbf{References}}

\addcontentsline{toc}{chapter}{APPENDICES} 
\appendix

%


%
\bibliographystyle{plain}
\ifthenelse{\boolean{PrintVersion}}{
\cleardoublepage 
}{
\clearpage       
}
\phantomsection  

\bibliography{MBZUAI-main}

\chapter*{APPENDICES}
\chapter{Papers and Implementations}

\section{Chapter 3 Implementation}
\textbf{Automatic Segmentation of Head and Neck Tumor: How Powerful Transformers Are?~\cite{sobirov2022automatic} 
}

Author contributions: Data Analysis and Preprocessing, Modeling and Implementation, Paper Writing.

PyTorch implementation can be found at: \\\url{https://github.com/ikboljon/hecktor_midl_unetr}.

The publication can be found at:
\\\url{https://proceedings.mlr.press/v172/sobirov22a.html}.

\section{Chapter 4 Implementation}
\textbf{Segmentation with Super Images: A New 2D Perspective on 3D Medical Image Analysis~\cite{sobirov2022segmentation}: 
}\\ 

Author contributions: Data Analysis and Preprocessing, Modeling and Implementation, Paper Writing.

The publication can be found at:
\\\url{https://arxiv.org/pdf/2205.02847.pdf}.

\section{Chapter 5 Implementation}
\textbf{An Ensemble Approach for Patient Prognosis of Head and Neck Tumor Using Multimodal Data~\cite{saeed2021ensemble} }

Author contributions: Data Analysis and Preprocessing, Paper Writing.

PyTorch implementation can be found at: \\\url{https://github.com/numanai/BioMedIA-Hecktor2021}.

The publication can be found at:
\\\url{https://link.springer.com/chapter/10.1007/978-3-030-98253-9_26}.

\textbf{TMSS: An End-to-End Transformer-based Multimodal Network for Segmentation and Survival Prediction~\cite{saeed2022tmss}: }\\ 

Author contributions: Data Analysis and Preprocessing, Modeling and Implementation, Paper Writing.

PyTorch implementation can be found at \\\url{https://github.com/ikboljon/tmss_miccai}.

The publication can be found at:
\\\url{https://link.springer.com/chapter/10.1007/978-3-031-16449-1_31}.

\end{document}